\documentclass[preprint,12pt]{elsarticle}
\usepackage{booktabs,multirow,soul, amsfonts, graphics, graphicx, amsmath, float, url}
\usepackage[ruled]{algorithm2e} 
\usepackage[a4paper, margin=1in]{geometry}
\usepackage{caption}
\usepackage{subfigure}
\usepackage{wrapfig}
\usepackage{xcolor}
\usepackage{makecell}
\usepackage{lineno}
\journal{IJHCS}
\begin{document}
\begin{frontmatter}
\title{Foraging-based Optimization of Menu Systems}



\author{Niraj Ramesh Dayama, Morteza Shiripour, Antti Oulasvirta}
\address{Aalto University, Finland}

\author{Evgeny Ivanko}
\address{Institute of Mathematics and Mechanics, Ural Branch, RAS \& Ural Federal Univ., Russia}

\author{Andreas Karrenbauer}
\address{Max Planck Institute for Informatics, Germany}

\begin{abstract}
The problem of computational design for menu systems has been addressed in some specific cases such as the linear menu (list). The classical approach has been to model this problem as an assignment task, where commands are assigned to menu positions while optimizing for users' selection performance and grouping of associated items. However, we show that this approach fails with larger, hierarchically organized menus because it does not take into account the ways in which users navigate hierarchical structures. 
This paper addresses the computational menu design problem by presenting a novel integer programming formulation that yields usable, well-ordered command hierarchies from a single model. First, it introduces a novel objective function based on information foraging theory, which minimizes navigation time in a hierarchical structure. Second, it models the hierarchical menu design problem as a combination of the exact set covering problem and the assignment problem, organizing commands into ordered groups of ordered groups. The approach is efficient for large, representative instances of the problem. In a controlled usability evaluation, the performance of computationally designed menus was $\sim 25\%$ faster to use than existing commercial designs. We discuss applications of this approach for personalization and adaptation.
\end{abstract}


\end{frontmatter}



\section{Introduction}\label{Introduction}\emph{Menu systems} are among the most prevalent user interfaces, 
offering a compact, extensible, and familiar means to access functionality.
Some popular menu techniques are known as linear, tabbed, hierarchical,
cascading, context, drop-down, ribbon, and toolbar menus. 
Designers frequently design menus, but their design remain challenging \cite{bailly2017visual}. 
Multiple objectives must be addressed,
including speed and accuracy of selection, learnability,
satisfaction, efficacy, suitability for different devices, and accessibility \citep{bailly2017visual}. 
Also the involved design spaces can be large.
%
%
Professional software, such as for photo-editing or 3D modeling, involve menus comprising of in excess of fifty commands.
It is not surprising that professional designers report menu design being ''very difficult'' and having to resort to trial and error \cite{bailly2013menuoptimizer}.

This paper contributes to algorithmic methods for generating and refining menu systems.
Our goals are (1) to improve the quality of generated menus and (2) support a larger number of commands (over 20 and up to 100) than previous research. 
Generally, larger menu systems need to utilize some type of \emph{hierarchical organization}, 
achieved by techniques such as tabbing, groups, folding, cascades, or sub-menus. 
Some promising advances notwithstanding, 
computational design of such hierarchical menu systems is still an unsolved problem. 
While there has been sustained research interest since the 1980s
\citep{ahlstrom2005modeling,
bailly2014model,
brumby2004good,
byrne1999eye,
mcdonald1983searching,
mehlenbacher1989finding, norman1991psychology, Paap1997533, sears1994split}, no
method has been offered that can automatically generate demonstrably usable and well-structured menus with a larger number of commands.
Professional designers would appreciate computational support on the matter \cite{bailly2013menuoptimizer}. 

Any algorithm for menu design will need to represent essential aspects of human behavior to be successful. 
Two challenges stand out: (1) the size of the search space and (2) lack of valid but computationally efficient evaluative (objective) functions. 
Firstly, the search spaces involved in menu design are exceedingly large:
$n$ commands can be organized into a linear menu in $n!$ unique ways and into a
hierarchical menu in an exponentially larger number of unique ways. 
If we consider multiple different tabs and also potential sub-groups within tabs, the space explodes further. 
Standard software applications commonly comprise dozens of
commands; professional applications may extend to hundreds of commands.  
The second issue, evaluative functions (objective functions), is even more challenging: the relevant literature has not yet identified any effective evaluative function that captures essential human factors mathematically. A well-known design objective -- given by Fitts' law -- characterizes the efficiency of selecting a command with a pointing device. Using Fitts' law leads to placing frequently accessed commands closer to the top of the menu \cite{bailly2013menuoptimizer}. Another objective investigated is related to grouping of items: 
placing associated commands near each other can make it easier to find them \cite{bailly2013menuoptimizer}. 
The association here can be based on distributional semantics (e.g., pairwise word associations) or on statistical co-occurrence in other menu designs.
Another factor in good menu design is the perceived balance between depth (tree) and breadth \cite{norman1991psychology}.
User expectations are a fourth consideration: Users may have preconceived notions, formed through exposure to prior
designs, of where in the menu certain commands should be found.
For instance, \emph{About} and \emph{Help} may be expected in the last tab.
To effectively apply computational methods for hierarchical menu design, a robust mathematical model and problem definition are needed that encompasses such considerations and yet allows efficient algorithmic solutions.

This paper presents a novel combinatorial optimization approach to the design of menu systems. 
It describes a mixed-integer  programming (MIP) formulation to handle
realistic-sized task instances.
It contributes a mathematical formulation of the menu design problem that (1) captures essential human aspects of menu navigation and (2) the decision problem in an efficient manner.
It produces well-structured and usable menu
designs when input data is provided for: (a) frequency of usage of individual commands and (b) mutual (semantic) association metrics for any pair of commands.
While previous research has mostly resorted to meta-heuristic techniques -- which are often based on randomization -- our
MIP approach guarantees optimality and provides mathematical estimates for bounds indicating the quality of a solution.
For any candidate solution, it is possible to compute bounds that indicate how far the current design is from the global optimum.

Two technical contributions are made.
The first lies in a new representative model of hierarchical menus.
Previous approaches used an assignment-based formulation \citep{bailly2013menuoptimizer}.
Over several studies of objective functions, 
we discovered that assignment alone does not sufficiently represent the organization of individual elements in larger entities such as groups or tabs. 
In particular, it leads to frequency-ordered groupings, at times ignoring how well the command placed at the top represents the rest of the menu. Hence, the topmost items are not necessarily semantically indicative of what the menu contains. 
In contrast, our formulation assigns each command to a group and then to a tab while also organizing (ordering) these for faster access.
In other words, both position (assignment) and grouping are naturally addressed in this new formulation,
unlike in previous work that only considered assignment.
Moreover, both assignment and grouping can be handled with a single objective (foraging),
which eliminates the need to set calibration weights.
To capture this critical aspect of how users navigate menus, we posit the design problem as a variant of \emph{the exact set covering problem}. Formally, the set covering problem is defined as follows: Given a finite
set $\mathbb{S}$ and a list of some (not necessarily all) subsets of $\mathbb{S}$, the intent is to find the minimal  sub-collection of disjoint sets such that all elements of $\mathbb{S}$ are covered exactly once. This covering problem precisely captures our intention of organizing the given menu commands into disjoint groups. 
The new objective yields organized \textit{groups of groups} with clear inter-group boundaries.  
This avoids the need to compute group boundaries \emph{post hoc} heuristically as in previous work using the assignment-based approach \citep{bailly2013menuoptimizer}. 

The second contribution is a new evaluative function based on \emph{information foraging theory} (IFT) \citep{fu2007snif,pirolli2007information}.
Previous literature focused on 
minimization of selection time \citep{bailly2017visual,ahlstrom2005modeling,card1980keystroke} and maximization of associativity among commands \cite{bailly2013menuoptimizer}. 
For example, \emph{MenuOptimizer} used Fitts' law and a statistical consistency metric measuring structural similarity of assignments to other menus \citep{bailly2013menuoptimizer}. 
Neither component specifies how the grouping of elements affects user behavior. 
Our contribution is a mathematically efficient formulation of IFT, which is made feasible for existing mixed integer programming solvers.
The new IFT-based objective enables assessing \emph{search performance} in the case of groups of groups,
which in our case are command groups (with separators between them) organized into tabs. 
Earlier work with IFT used it for modeling how users choose link panels \citep{10.1007/3-540-44963-9_8}.

In the case of hierarchically organized menus, 
it offers a quantitative model of a rational but time-limited agent navigating a hierarchy composed of patches.
The agent decides whether to continue exploring the current set of commands (patch) or instead abandon/skip this set in favor of the next.
Intuitively, when used in an optimizer,
it evaluates and minimizes also the time \emph{wasted} by the user in the irrelevant parts of the menu. 
This results in positioning of semantically indicative items toward the top of the menu.
In practice, this is achieved by three means:
(1) The optimizer forms groups that enable users to quickly guess whether the intended
command is present or not. 
(2) Secondly, it inherently avoids too high a number of
groups/tabs. 
(3) Finally, it avoids placing unrelated commands (\textit{loners}) in
groups with poor association. 

The convergence of these two ideas -- the exact set covering formulation with the IFT-based evaluative function -- 
yields balanced and well-structured menus.
Since the decision to include a command in a group and tab is modeled explicitly in the problem, no \emph{post hoc} steps are needed to segment the outputs.
The menus thus produced consist of a few tabs that, in turn, are made up of relatively large and well-organized groups.
They also appear better for comprehension in terms of their structure than were results of previous work,
also because the lead elements are semantically more indicative.  
It is easier to recognize the idea of a tab or group and act accordingly -- that is, dismiss it or, if it is relevant for the goal at hand, zoom in.
Also, the MIP implementation does not require extensive computational effort.
The resulting formulation can deal with problem instances of 50 commands within a few hours of computational effort. 
While larger instances of, say, 100 commands, take about two weeks' computational effort on a regular computer,
in the context of a large design project.
Presently it does not produce labels for the higher-level groups it has created, such as for tabs in a tabbed menu.

A designer can use the outputs to explore the design space  or fine-tune an existing design \cite{oulasvirta2020combinatorial}.
To use the optimizer, a design task (instance) must be defined. 
The inputs are (1) a list of command frequencies and (2) a matrix of pair-wise association scores (0--100).
These can be given by the designer or obtained in a data-driven fashion. 
Access frequencies can be learned for example from click data, or estimated using click models.
Word embeddings can be used to estimate pairwise association scores. 
Alternatively, when available, word association norms (e.g., based on WordNet) could be used.
Also, co-occurrence of commands can be learned from existing menu designs as done in previous work \cite{bailly2013menuoptimizer}.
Association scores are relatively easy to provide manually too. 
Because the association matrix is sparse (only a few cliques of commands have meaningful relationships and the rest can be skipped), filling in the matrix does not take too much time even for larger problem sizes.
In our evaluation cases, a student could define an association matrix (for about 50 commands) within an hour.

To critically assess the approach, we report on a controlled comparison between optimized and commercial designs (Adobe Reader, Microsoft Notepad, and Mozilla Firefox). 
The results of our new approach demonstrates that users could work 25\% faster with our optimized menus compared to the existing designs.
The new approach is able to produce high-quality designs.
We selected Adobe Reader, Microsoft Notepad, and Mozilla Firefox because these three applications represent a cross section of well-established commercial software systems; the commands and context of these systems should be comprehensible to an average user without separate training or explanations.  We point out the fact that this comparison baseline does not inherently include multi-level sub-menus; this matches our approach because we also do not address multi-level menus in the current paper. Further, the remainder of this paper intentionally does not discuss the naming of the tabs. Assignment of names involves natural language processing problems which is outside the scope of this work. 

%
The rest the paper is organized as follows:
We first present a succinct review of the literature related to algorithmic menu design and related areas in operations research. 
After this, we define the design problem rigorously and use the definition to inform a 
``classical'' MIP formulation that replicates the objective function used in previous work using Fitts' law and associativity as objectives. 
We then extend this formulation to utilize the
IFT-based approach within the MIP formulation, 
after which we discuss applications also in personalization and adaptation of menus. 
We finally present a user study comparing
our algorithms' results with commercial baseline designs.
Finally, data from the controlled user evaluation is presented.

\section{Problem Definition} \label{Definition of the Problem}This paper addresses the problem of finding an optimal layout of commands in a hierarchical menu structure. 
The instance of a hierarchical menu examined here is the popular tabbed menu system in which commands are organized into groups, which are arranged into tabs. 
The most common menu types (linear menus, ribbon menus, etc.) are special cases of this general formulation that can be modeled by changing costs in the evaluative function. Our underlying objective with this paper -- as reflected in the evaluative functions we explore -- is to minimize the overall time consumed by the user in selecting commands.

\begin{figure}[!htbp] 
\centering
\scalebox{0.33}
{\includegraphics{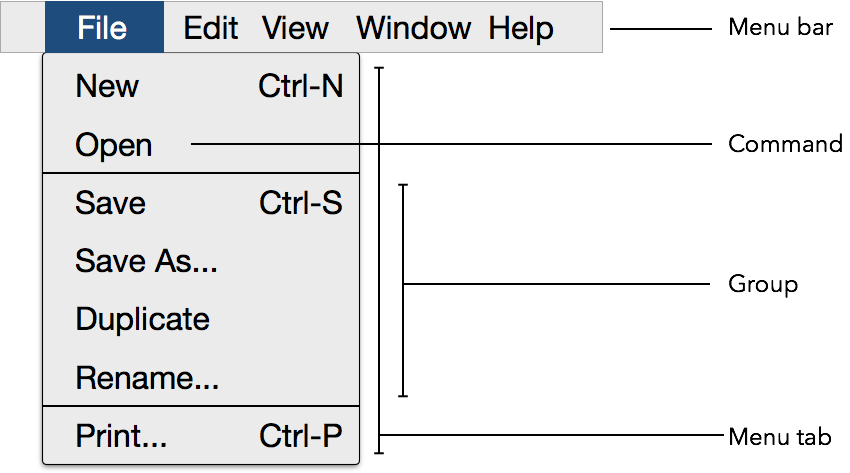}}
\caption{Illustration of key terms utilized in this paper. The design task we examine is how to assign commands into groups and tabs accessible from a menu bar.}
\label{terminology2.png}
\end{figure}

The key terms utilized in our definition are illustrated in Figure \ref{terminology2.png}. 
We call a menu item a \textit{Command}. A clearly demarcated set of commands (separated from other groups by a physical boundary) is termed a \textit{Group}. Multiple groups arranged in vertically aligned structures constitute a \textit{Menu Tab} (or tab). 
Individual tabs have their specific identifier text or titles; all of these titles collectively constitute the horizontally aligned structure denoted as a \textit{menu bar}. For clear disambiguation, we note that this paper addresses single-level hierarchy within menus; this means that one \textit{Menu Tab} in the menu bar will involve one vertically aligned structure of commands that must all be visible or hidden as an indivisible set. This paper does not support any further internal roll-over within the \textit{Menu Tab}. While multiple groups are permitted within any \textit{Menu Tab}, they must all be visible or hidden together.

Given $n$ commands, the problem first requires these commands to be organized as ordered sets, each representing one group.
Then, these groups are themselves organized into ordered sets that form individual tabs. Lastly, the tabs, in turn, are ordered in terms of the labels to be shown in the menu bar. This means that the overall layout is a problem of reorganizing the unordered set $N$ into an ordered set of all $n$ commands. 

The $n$ commands are characterized by frequency of use $\mathbb{F}$,
associations $\mathbb{A}$, and (optionally) location preferences $\mathbb{L}$. We will denote the set of tabs as $T$, with individual tabs identified by index $\tau$. All groups (irrespective of their tabs) are assumed to constitute an unordered set of groups $C$. Individual  groups will be identified by indices $c,\bar{c}$. We note that any tab $\tau$ is itself an ordered set of some groups. Further, any group
$c \in C$ is also an ordered set of some commands. Finally, we use indices $i,j$ to denote commands. Hence, the frequency of usage of command $i$ is $\mathbb{F}_i$. The association between commands $i$ and $j$ is denoted as $\mathbb{A}_{ij}$. So, the menu design problem requires computation of an ordered set $T$ of ordered sets of groups of ordered
commands, such that the objective function (defined via values of $\mathbb{F}$, $\mathbb{A}$, and optionally $\mathbb{L}$) is optimized.

\subsection{Objective function: Costs to minimize}
The problem definition discussed above did not explicitly provide the objective function. Traditionally, menu design has involved an  objective function that is a weighted combination of two factors: (1) Time required to access individual commands (2) Mutual association of commands placed within any group  (see, e.g.,  \cite{norman1991psychology,cockburn2007predictive,goubko2010automated,danilenko2013semantic,bailly2014model,chen2015emergence,bailly2017visual,oulasvirta2018combinatorial}). This objective function helps in two ways. First, the time required by any user -- who knows the location of a required command -- is exactly captured using Fitts' law. So, the efficacy of an expert user (well conversant with the concerned application) is explicitly captured by the objective term. This ensures that the resulting menu system is indeed fast for usage by a well-experienced user. Secondly, the mutual association term helps a novice user in searching for any command quickly. Consider that a user is looking for the Save-As command. While exploring the menu, the user had encountered the Save command in the first (leftmost) tab. The user remembers this location and also associates the required Save-As command with the known Save command. So, the user is highly likely to search for the Save-As command in vicinity of the known Save command. The second objective term -- representing the mutual association of commands -- ensures that logically interrelated commands are collocated in close vicinity to each other.  

In this paper, we do provide a classical formulation using this traditional objective function in Section \ref{Minimal Representative Formulation}. But then, we also propose a novel objective function based on the Information foraging theory in Section \ref{The Information Foraging Approach}. For either formulation, our concept of cost encapsulates the efficiency of a broad set of users: While addressing the speed-up for expert users who know the application well, we also wish to assist the exploratory efforts of novice users who are often searching for required commands with only a vague idea of requisite command names.

\subsection{Scope}
There exists a wide variety of interaction techniques for menus and menu-like paradigms for command selection \cite{bailly2017visual}. 
This paper specifically focuses on non-adaptive menus with hierarchical structures. 
The positions of commands are not assumed to change during interaction.
We target designs wherein commands are organized by groups and tabs. The objective function does not consider all aspects of menu design: selection of labels or shortcuts, item length, etc. For the purposes of this paper, navigation occurs by selection of a tab and then a command within the tab. In this instance of hierarchical menu systems, only a limited subset of a menu tree can be visible at a time. While it is possible to extend the techniques in this paper to cover some other types of hierarchical menus, that discussion is beyond the scope of the paper.


\section{Related Work} \label{Related Work}Our work builds on results in four areas of related work:
(i) modeling of search performance for menus,
(ii) meta-heuristic techniques,
(iii) integer program (IP) approaches based on the assignment problem, and
(iv) facility layout and next release problems in operations research and software engineering.

\subsection{Search performance and predictive models}

Human factors in selection performance have been studied extensively.
There is evidence of the following empirical effects:
\begin{enumerate}
  \item Shorter menus are faster to use \citep{anderson1997act,byrne1999eye,nilsen1996perceptual};
  \item Targets that are closer to the top are faster to select \citep{byrne1999eye,cockburn2007predictive};
  \item Linear menus with grouping (``semantic'' or ``systematic'' menus) are
  much faster than random or alphabetical ones \citep{danilenko2013semantic,mcdonald1983searching,hornof1997cognitive,mehlenbacher1989finding};
  \item It is faster to select a target that is present in the menu compared to determining that the desired target is not present in the menu \citep{bailly2014model};
  \item Users get faster with practice, and this positive effect influences
  other (non-target) items in the sub-menu also \citep{cockburn2007predictive}; and
  \item Users often fixate on one of the first three items \citep{byrne2001act}.
\end{enumerate}
Prior mathematical models \citep{bailly2014model,cockburn2009predictive}, typically using non-linear regression, capture some of these effects.
Cognitive simulations such as EPIC \citep{hornof1997cognitive}, ACT-R \citep{byrne2001act},
and computational rationality \citep{chen2015emergence} capture more effects
but are very computationally expensive.

It has been found that users utilize various \emph{search strategies} in menu navigation:
\begin{enumerate}
  \item \emph{Directed search} uses memory of element locations learned through
  practice for guessing where to search \citep{bailly2014model};
  \item \emph{Serial search} progresses downward from the first or topmost command, after which items are sequentially examined one at a time \citep{byrne1999eye};
  \item \emph{Random search} selects an arbitrary position within the menu for fixation \citep{byrne1999eye,hornof1997cognitive}; and
  \item \emph{Visually guided search} is based on sampling of visual landmarks
  such as the end of the menu or labels sharing visual features with the target \citep{bailly2014model,chen2015emergence}.
\end{enumerate}

No previous mathematical model has precisely predicted the search behavior in a hierarchical menu. 
Sears and Shneiderman \cite{sears1994split} presented a model for \emph{split menus},
where frequently used items are placed at the top of a linear menu, in their own
group. Their model assumes that search time for frequently selected items
is a logarithmic function of distance from the top and for low-frequency items is a linear one.
Lee and MacGregor \cite{lee1985minimizing} proposed that selection time follows the number of
pages needed to access retrieval of a given item, number of items per page, time
needed to assess one item, keystroking time, and system response time.
These models do not account for the effects of any other design
factor than the number of items on a page or in a group.
Bailly and colleagues \cite{bailly2014model} introduced a free parameter to their model, indicating
which of two search strategies is used.
A gaze pattern was found wherein experienced users fixate on the first items in
sub-groups and then either drill down or not.
However, the pattern was not captured well by the model, and the authors identified
this as a target for further improvement. A recently presented model \citep{chen2015emergence} suggests that the optimal search strategy adapts to the semantic organization of a linear menu, allowing users to gaze more directly at relevant sub-groups. This model relies on reinforcement learning, and it has been tested for predicting the effect of menu organisation and menu length. on task completion time and eye movements.

With this paper, we investigate IFT's suitability for a computationally efficient model capturing a key aspect of hierarchy-related decision-making: the decision to keep reading a sub-menu (tab or group) versus jump to the next candidate.

\subsection{Menu optimization using meta-heuristic techniques}

Most prior work \cite{Troiano2008242,Troiano2014433} on menu optimization has used a meta-heuristic technique. Meta-heuristic techniques do not make explicit assumptions about the objective function; rather, they consider it as an oracle that tells them the objective value of a given candidate. 
As a result, they can work with any objective function, including non-linear functions and even simulator models.
In contrast to exact methods such as integer programming,
meta-heuristic techniques cannot offer guarantees that the best design found is optimal.
Moreover, meta-heuristic techniques, such as simulated annealing and genetic algorithms, include many hyperparameters,
the tuning of these hyperparameters may affect their performance.

Troiano et al. \cite{troiano2016searching} used an index of accessibility and a user preference
indicator to define objectives in a genetic algorithm (GA) solver that operates from the number of
children of the item and depth of the menu hierarchy. Matsui and Yamada \cite{matsui2008optimizing}
explored objective functions to address selection time
that consist of search and pointing, a penalty term for functional dissimilarity
with other designs, and a menu granularity regularizer.
As the authors noted, the approach is brittle, because results can change
dramatically with small adjustments to objective weights.
The state-of-the-art approach at the moment is MenuOptimizer
\citep{bailly2013menuoptimizer}.
It uses a weighted sum from a selection time model (SDP
\citep{cockburn2009predictive}) and, as in Troiano et al.'s work, a structural metric.
A weakness due to the use of the latter is that it produces results that
are driven to be similar with previous designs. Further, the quality of the resulting solution changes with every execution.
However, reasonable results can be produced quickly, within 5--15 minutes even on commodity hardware.
The design of hierarchical menu systems has been restricted to the realm of meta-heuristic techniques, in areas such as simulated annealing \citep{matsui2008optimizing},
genetic algorithms \citep{golovine2010evolving,troiano2016searching}, and ant colony optimization \citep{bailly2013menuoptimizer}.

Heuristic constructive approaches to optimization generate candidate designs using some heuristic that has been found to work in the domain. 
In menu optimization, this approach has been explored as a
combination of exhaustive search and ``drill-down'' \cite{danilenko2013semantic,goubko2010automated,goubko2012mathematical}.
That is, it initially explores the most relevant solution
directions, applying a breadth-first paradigm, then chooses a few suitable candidates for in-depth 
inspection. 
The methods presented thus far assume, however, that
potential groupings of commands are stated \emph{a priori}. 
This is a limitation, 
because the grouping predetermines the optimal solution. 
In this, the task becomes harder for the designer and relatively easier for the optimizer,
which need only solve the ordering problem.
Often, the grouping is a defining part of the problem at hand.

The method described in this paper requires only associations among frequent elements, without any pre-grouping of elements. However, if the \emph{a priori} grouping were to be already available, the methods in this paper can make efficient use of this knowledge: Association values can be expressed using a binary mapping (value 1 when concerned commands are in same group and 0 otherwise).  

\subsection{Exact methods: Menu optimization as an assignment problem}

Unlike meta-heuristic techniques, exact methods are guaranteed to find the optimal solution in finite time.
However, the time required may be an exponential function of the problem size (most interesting problems are NP-hard).
The simplest exact method is \emph{explicit enumeration}, wherein the objective value of each element in the solution space is evaluated and the current best solution (the so-called \emph{incumbent}) is updated. 
In contrast, \emph{implicit enumeration} makes use of relaxations that can be solved efficiently.
One very popular form of implicit enumeration, which we use here, is \emph{Branch \& Bound}. This is one of the standard methods for solving Integer (Linear) Programs.

\emph{Keyboard layout design} was defined in the 1970s as a quadratic assignment problem wherein the goal is to minimize the average time for movement between letters assigned to buttons. 
Relaxations to the quadratic assignment problem (QAP) formulation have reduced solution times with integer programming (IP) solvers
to a permissible range even with realistic data
\citep{karrenbauer2014improvements}.

The simplest application of this approach to menu design arrives at solutions for a \emph{linear assignment task}
\citep{oulasvirta2018combinatorial}; the formulation can be extended to
full menu systems using hierarchical structures. 
Yet, designs created using assignment alone display systemic shortcomings. In particular, items are not assigned to groups, which precludes hierarchy. In addition, the most important items in each group are placed in the center (because of the associativity term). 
Moreover, though the design task has been formulated in integer programming terms, the problem has to be finally solved by means of meta-heuristic techniques due to 
the expensive nature of the evaluative functions required.

\subsection{Selection problems in diet planning and service design}
\emph{Menu planning} (or diet planning) for the restaurant industry was defined as a linear programming
problem in 1964 \citep{balintfy1964menu} and has received sustained interest since
\citep{lancaster1992history}.
The task here is to select food combinations that meet dietary, gastronomic, and
production objectives.
The \emph{diet problem} is a variant of menu planning for an individual or group
\citep{dantzig1990diet}.

While food-menu planning is not relevant for the design of menus in computing systems,
the selection of functionality is.
Functionality selected for an application or service must be accessed via a menu
system.
What is known in software engineering as \emph{the next release problem} refers to selection
of new software features \citep{bagnall2001next} that accounts for
user preferences, dependencies among functions, and costs.
In a recent paper \citep{oulasvirta2017computational}, integer linear programming
 (ILP) was presented for the selection of functionality in
interaction and service design.
Here, we do not discuss the problem of how to select the menu
commands; we assume the set $N$ to be known.

\subsection{Facility layout problem}
Facility layout problem (FLP) is a class of combinatorial decision-support
system that deals with the location (placement) of facilities for factories \citep{DRIRA2007255}.
Facility location translates to the search for an optimal arrangement of
non-overlapping indivisible entities within a given structure. The objective
measure to be minimized is defined in terms of the weighted  distance between
the centroids of the entities being positioned.
Within the broad FLP field, 
closest to menu design is the multi-row facility
layout problems, or \citep{ANJOS20171}, a variant that
allows for a layout with multiple rows (in a predetermined number) to which the entities
can be assigned. The entities all have the same size (with height equal to the common
row height), distances between adjacent rows are equal, and entities can be
assigned to any row in general.
We note a  parallel between FLPs and our problem.
The arrangement of commands into unique rows within tabs
matches the multi-row facility layout problem (MRFLP). 
The literature discusses mixed-integer programming
and also semi-definite optimization approaches for MRFLP (see
Gen \& Cheng \citep{MRFLPBook}). Regrettably, both approaches involve a highly
non-linear objective function, which adversely affects computational
performance. We cannot expect these prior formulations to address problems
with sizes beyond 15--20 menu commands and with additional complicating design
considerations.
Moreover, their objectives are different and not directly to menu use, which emphasizes comprehensibility and fast access.

\section{Reformulating the Design Task: A Minimal Representative Formulation} \label{Minimal Representative Formulation}The first part of our method is a flexible integer programming
formulation for key decisions in designing a hierarchical menu structure. 
Intuitively the problem is defined as ordered sets at multiple levels. 
We pursue a formulation that is compact enough to produce results with limited computational effort. 
It is flexible also in the sense that it can support evaluative functions of numerous types. 
We designate the formulation as a \emph{minimal representative formulation} (MRF). 

The primary intention behind the MRF is to represent the structure of the \emph{hierarchical} menu design problem, expanding from the assignment problem as dealt with by previous work. 
This lets us plug-in various evaluative functions and thereby benchmark existing approaches to menu optimization.
The formulation relies on principles utilized in the relevant technical literature but not previously explored for menu optimization.
These principles have made it possible to overcome some limitations of the assignment-based approach in the hierarchical case. 
In particular, we can now express an evaluative function that refers to a user's navigation behavior both at the level of commands and at the level of sets of commands.
We exploit this to construct an information-foraging-based evaluative function that is more natural for hierarchical menu systems than the Fitts' law and associativity matrix approach (see the next section).
The MRF is developed here as a mixed-integer linear program. We try out two evaluative functions to compare their results. The final outcome from the MRF is expected to represent the best possible results that can be obtained via techniques and approaches from prior literature.

Technically, the MRF employs \emph{decision variables} that map specific commands to individual
locations (row numbers within tabs).
To address the set-cover-related consideration of
determining the intra-tab grouping of commands, we define decision
variables to represent the number of groups and also the constitution of any
individual group. Finally, to
ensure the integrity of groups and row numbers (i.e., avoid ``holes'' between row numbers),
we use a general precedence variable that avoids non-linear terms while
ensuring the expected computational performance. This results in a compact model with
fewer variables and constraints.\\
\newline
\textbf{Decision variables:}
\begin{align}
&X_{i}^{c} = 
\left\{
\begin{array}{lcl}
{1} & {\dots} & \mbox{if command $i$ is placed in group $c$}\\
{0} & {\dots} & \mbox{otherwise}
\end{array}
\right.\notag\\
&Y_{i}^{\tau} = 
\left\{
\begin{array}{lcl}
{1} & {\dots} & \mbox{if command $i$ is placed on tab $\tau$}\\
{0} & {\dots} & \mbox{otherwise}
\end{array}
\right.\notag\\
&Q^{c\tau} = 
\left\{
\begin{array}{lcl}
{1} & {\dots} & \mbox{if group $c$ is placed on tab $\tau$}\\
{0} & {\dots} & \mbox{otherwise}
\end{array}
\right.\notag\\
&Z_{ij} = 
\left\{
\begin{array}{lcl}
{1} & {\dots} & \mbox{if commands $i,j$ are placed in the same group}\\
{0} & {\dots} & \mbox{if commands $i,j$ are placed in different groups}
\end{array}
\right.\notag\\
&W_{ij} = 
\left\{
\begin{array}{lcl}
{1} & {\dots} & \mbox{if commands $i,j$ are placed on the same tab}\\
{0} & {\dots} & \mbox{if commands $i,j$ are placed on different tabs}
\end{array}
\right.\notag\\ 
&R_{i}^{r} = 
\left\{
\begin{array}{lcl}
{1} & {\dots} & \mbox{if command $i$ is placed on the $r^{th}$ row of some tab}\\  
{0} & {\dots} & \mbox{otherwise}
\end{array}
\right.\notag \\
&t_i = \mbox{The time required to reach command\ $i$ as indicated by Fitts' law}\notag 
\end{align}
\begin{align}
&S^{c\bar{c}} = 
\left\{
\begin{array}{lcl}
{1} & {\dots} & \mbox{if group $c$ immediately precedes group $\bar{c}$ on
some tab}\\
{0} & {\dots} & \mbox{otherwise}
\end{array}
\right.\notag\\
&\mathbb{S}^{c} = 
\left\{
\begin{array}{lcl}
{1} & {\dots} & \mbox{if group $c$ is the first (topmost) group on some
tab}\\
{0} & {\dots} & \mbox{otherwise}
\end{array}
\right.\notag\\ 
&\xi^{c} = 
\left\{
\begin{array}{lcl}
{1} & {\dots} & \mbox{if group $c$ is used (has a non-zero number of commands)}\\
{0} & {\dots} & \mbox{otherwise (empty, with no commands)}
\end{array}
\right.\notag \\
&\beta^{\tau} = 
\left\{
\begin{array}{lcl}
{1} & {\dots} & \mbox{if tab $\tau$ is used (has a non-zero number of groups)}\\
{0} & {\dots} & \mbox{otherwise (empty, with no groups or commands)}
\end{array}
\right.\notag \\
&\Theta^{c\bar{c}} = 
\left\{
\begin{array}{lcl}
{1} & {\dots} & \mbox{if group $c$ is placed anywhere before group $\bar{c}$
on some tab}\\
{0} & {\dots} & \mbox{otherwise}
\end{array}
\right.\notag  \\
&P^c = \mbox{Starting position (row number) of group $c$ within its tab}\notag 
\end{align}
Decision variables $X,Y,Q$ define the unordered structure of the groups and
tabs. Decision variables $S$ and $\Theta$ offer alternative ways to
enforce the relative ordering of groups. The absolute
positioning of commands is provided by $R$. The variables $W,Z,\xi, \beta$ ensure the sanctity of the overall mathematical model. By "sanctity", we imply that these variable are intended to ensure that the decisions enforced by the other variables are in synchronization with each other. In the absence of $W,Z,\xi, \beta$, the results implied by the other variables can be infeasible or non-unique.   

The decision variable $\Theta$ requires more explanation. Classical mixed-integer programming
formulations handle sequencing of elements by using an immediate precedence
variable; this variable typically specifies that some element $i$
\textit{immediately} precedes some element $j$. The immediate precedence
variable inherently dictates the relative ordering and also the collocation;
there must not be any element $k$ between $i$ and $j$. In contrast,
$\Theta$ specifies the relative order alone and not
the collocation. Hence, one or more elements may be present between $i$ and
$j$. This general precedence variable provides several logical distinctions from
the immediate precedence approach, and it will be required for row numbering. The full set of constraints applied to these decision variables is covered in the Appendix (Subsection \ref{MRFConstraints}). 

\subsection{Example application: An evaluative function based on previous work}
\label{Classical Objective Function for MRF}
A key benefit of the MRF is that it can be used with a broad range of evaluative functions. 
Here, we replicate the ``two-fold-objective'' function of Bailly et al. \citep{bailly2013menuoptimizer}, which balances (i) the time required to reach the commands (weighted by the frequency of use) against (ii)
the association of commands placed in a specific group or on a certain tab.
\begin{equation}
\mbox{Maximize: }\ 
	\sum_{i\in N} \sum_{j\in N} \mathbb{A}_{ij} \left( \lambda_c {Z}_{ij}
	+ \lambda_m {W}_{ij} \right) -\lambda_f \sum_{i\in N} \mathbb{F}_i t_i  \label{MRF_Objective}
\end{equation}
The parameters $\lambda_f$, $\lambda_c$, and $\lambda_m$ are the relative weightages for access time (from Fitts' law), intra-group associations, and inter-group (intra-tab)
associations. We will discuss values for these weightages in Subsection \ref{Parameter values}. We designate this as the two-fold objective because we are capturing two different metrics here: the term in $\mathbb{A}$ implies the effort in guessing the location of a command; the term in $\mathbb{F}$ captures the time required to reach a particular command when the user already knows its location.  
The results obtained for this objective function are presented, in Section \ref{Results}, for comparison of the results to those obtained from the evaluative function based on the information foraging theory.

\section{The Information Foraging Approach} \label{The Information Foraging Approach} We develop a new evaluative function based on the information foraging theory (IFT) \cite{pirolli1999information}.
IFT models search behavior as utility-maximization in a patch world.
The theory is an application of the optimal foraging theory in biology, 
which describes the hunting and food search behavior of animals. 
An adaptive agent is assumed to change the patch as soon as the gain decreases to a level that it would make more sense to move elsewhere. 
Consider a wolf that has nearly exhausted the food in its current forest. 
Should it stay there, go to a nearby forest with rabbits (representing few calories per day), or travel a bit further to reach a different patch with deer (which are harder to catch but offer a greater gain)?
Application to menu interaction follows analogous reasoning. 
Just as in food foraging, the (information) ecology of a menu is \emph{patchy}; 
that is, information about the target is unevenly distributed to patches such as groups and tabs. 
Because some of the patches are not fully visible and are accessed only via higher nodes in the tree, 
the agent must decide what to attend under uncertainty. 
From what is locally visible (e.g., the first items in a group), 
the user must infer what the rest of the region may carry. 
Hence, the key to menu design is not how close an item is to the top, or to related items, but how economically the user can decide how well it represents the rest of the menu.

To model this kind of ``information scent'' \cite{pirolli1999information}, 
as is done in IFT applications in general, 
we assume that an agent's environment consists of patches indicative of a target to varying degrees and connected with distances (or time costs).
Each patch is associated with a gain: 
a function describing how quickly information is extracted when the user is in that patch.
In IFT, a non-linear (logarithmic) function is used to model gain.  
It is continuous and has the property of diminishing returns:
as more time is spent in a patch, less information becomes available,
and a rational forager moves to another patch. 
The theory further posits that the user must make a decision for every set of
commands attended. 
In the case of a menu, 
the user looks at the leading (top) commands in the set (e.g., under ``File'') and then makes a guess as to whether or not this set of commands may contain the desired command (e.g., ``Zoom In'').
If the user guesses that the current set should contain the required command, then the user
will investigate further by reading (exploring) within this particular set. 
Otherwise, he or she discards the current set and moves on to the next \textit{without really analyzing the content of the current set}. 

To make IFT amenable to mixed-integer programming, we have formulated a sample--discard--explore paradigm that allows avoiding non-linearities (e.g., logarithmic gain functions) but retains the essence of this foraging behavior.
Intuitively, the sample--discard--explore function captures four logical outcomes possible during search:
\begin{enumerate}
  \item \textbf{True positive}: The user guesses that the set contains the 
  target command, and it indeed contains that command. In this case, the cost during search 
  within the set is the time consumed (by Fitts' law) to scan the list and move the pointer over the
  required command in the set. 
  \item \textbf{True negative}: The user guesses that the current set does \emph{not} contain
  the required command, and the set indeed does not contain it.
  No further cost is incurred. 
  \item \textbf{False positive}: The user guesses that the set contains the
  required command, but it actually does not. The additional cost incurred for this set is the time consumed (under Fitts' law) to navigate all
  commands in the set. This cost is proportional to the size of the set.
  \item \textbf{False negative}: The user guesses that the set does not contain the
  required command, but in reality it does. Now the user must (fruitlessly) analyze all succeeding sets,
  such as subsequent groups on the tab. 
\end{enumerate}
We assume that the user begins the search by sequentially analyzing (sampling) the lead elements of every set encountered.
On the basis of the decision made to discard or explore any specific set, the user invests the corresponding effort for that set. This process repeats until a true positive (target) is reached.
This logic can be applied recursively at any level of a hierarchy where multiple options (sets) are available. In our application we assume two levels: Tab and Group.
The insight is that the total time expended in locating a specific command is the summation of time spent in the four possible scenarios, weighted by the probability of the user making the corresponding decision for the relevant set. 
To obtain an estimate for the entire menu structure generated by an optimizer, 
the estimated times are further weighted by the frequency of use of individual commands.

In addition to the search-related
time components, motor selection efforts must be considered. 
Consider the case where the user already knows or remembers that some
command $i$ is in group $c$ on tab $\tau$ at row number $r$. There
is no search effort at all, yet the user still takes some time to traverse
to row $r$ on tab $\tau$. As in previous work, this time is as computed from Fitts' law, and it
depends on $r$ and $\tau$ only. We address this time via the decision variable
$t_i$.

We also need to quantify the user's expectation of a specific group or tab featuring command $i$.
This expectation depends solely on
the user's current knowledge of the presence/absence of other commands (such as $j$) in
the group or on the tab. We quantify this expectation as follows:
\begin{enumerate}
  \item If the desired command ($i$) is the leading (topmost) element of any
  group, the expectation is 100\% for that group. 
  \item If leading element $j$ of any group has a very high
  score (above 80\%) for association with command $i$, then the expectation is
  100\% for that group. Conversely, a very low score (below 20\%) for association between $i$
  and $j$ leads to a negligible expectation. 
  \item For intermediate, unexceptional association score values, the
  expectation is scaled in proportion to the relative value of the association
  score with respect to the median one.
\end{enumerate}
Given a group $c$ and an element $i$, we can judge the expectation of
the presence of $i$ in $c$ by looking at lead element $j$ of group $c$. Hence,
the expectation of $i$'s presence depends solely on the association between $i$ and
$j$. We denote this expectation as $E^{ij}$. We note that $E^{ij}$ can be
computed in advance through a pre-processing step, so it can be treated as a known
parameter in the mixed-integer programming formulation. We use $E^{ij}$ to scale the efforts
in every group for every command.

\subsection{Mathematical formulation}
To develop an integer programming formulation based on IFT, we require decision
variables that can uniquely specify the solution (the resulting layout) while
enabling computation of the various efforts mentioned above. The specific decision variables (including computation of specific efforts) are explained below. We note here that the decision variables described in the previous section are required too, along with their constraints.

\begin{align}
&U_c^{j} = 
\left\{
\begin{array}{lcl}
{1} & {\dots} & \mbox{if command $j$ is the topmost (lead) element of group $c$}\\
{0} & {\dots} & \mbox{otherwise}
\end{array}
\right.\notag\\
&\Phi_i^c = \mbox{Total time / cost for command\ $i$ computed for group\
$c$}\notag\\
&\alpha_i^c = \mbox{True-positive time / cost for command\ $i$ computed for group\
$c$}\notag\\
&\sigma_i^c = \mbox{False-positive time / cost for command\ $i$ computed for
group\ $c$}\notag\\
&\delta_i^c = \mbox{False-negative time / cost for command\ $i$ computed for group\ $c$}\notag\\
&\Omega_i^\tau = \mbox{Penalty incurred for command $i$ if it is placed on (non-standard) tab\ $\tau$}\notag
\end{align}
The decision variable $U_c^{j}$ assists in locating the lead element in any group. Our IFT approach is based on this lead element, so its knowledge is critical. The variable $\Phi_i^c$ encapsulates the expected total time, cost or effort required to reach the command $i$ if it were to be placed in group $c$. The value of $\Phi_i^c$ will be a function involving the location of group $c$, its lead element and overall composition. This value of $\Phi_i^c$ is expressed in terms of the three possible navigations -- true-positive, false-positive and false-negative as covered by $\alpha_i^c, \sigma_i^c, \delta_i^c$ respectively.

The objective is to minimize weighted cost $\Phi$ (weighting is by frequency
of use) for the time taken to reach any command placed within any set. 
\[
\mbox{Minimize } \ \  
\sum_{i\in N} \mathbb{F}_i \left( \ \ 
\sum_{c\in C} \Phi_i^c + \sum_{\tau \in T}  \Omega_i^\tau
\ \ \right)
\]
such that: 
\begin{align}
&\Phi_i^c\ \ge\ \lambda_0 t_i\ +\ \lambda_1 \alpha_i^c + \lambda_2 \sigma_i^c + \lambda_3 \delta_i^c \ \dots \forall\ i\in N, c\in
C\\
\intertext{This constraint is the key to the IFT approach and requires more explanation. 
Here, $\Phi_i^c$ does not intend to capture the exact time spent by a specific user in finding a specific command during a specific single session of usage. Rather, it intends to encapsulate the weighted estimate of sum of searching effort and accessing effort for that command. The first term of this constraint represents the time (computed via Fitts' law) needed to navigate to command $i$ if its location is known in advance. But the wasted efforts from false positive or false negative -- and even the searching effort from true positive -- must also be counted while optimizing the location of $c$. The remaining terms indicate the relevant costs incurred in searching for command $i$ with respect to group $c$.
Hence, $\Phi_i^c$ is the summation of all concerned costs. The $\lambda$ values are weight factors to be specified by the designer.} 
&X_i^c \ge U_i^c \dots \forall i \in N, \forall c\\
&\sum_{i\in N} U_i^c = \xi^c \dots \forall c\\
\intertext{These constraints ensure that exactly one command is marked as the
lead element of every group. In addition, we augment Equation
\eqref{LowerLimitOnRowNum} to ensure that command $i$ marked as the lead
element of any group $c$ has its row number equal to $P^c$. Next, we look at
constraining the values of individual cost components.
To calculate an individual cost component, we take the scalar product of the
expectation value (probability of exploring the set) and the time expended in exploration of this set. For example, 
the expected expense of a false positive for command $i$ in group $c$ is as follows:} 
&\sigma_i^c \ge E^{ij}  \sum_{k\in N} X_k^c - \nabla (1+X_i^c-U_j^c) \dots \forall i,j \in N, \forall c\\
\intertext{Here, $\nabla$ is a suitably chosen sufficiently large constant. If the $\nabla$ related term is neglected, then the false-positive cost is computed in terms of the size of the group that was needlessly explored. The $\nabla$ related term voids the constraint if $j$ is not the lead element or if $i$ actually is present.  Thus, the constraint addresses the case of a false positive occurring when group $c$ is led by element $j$ and the user is exploring $c$ to search for $i$ when $i$ is not, in fact, present in $c$. Next, let us consider the case of a false negative for command $i$ in group $c$:}
&\delta_i^c \ge (1-E^{ij}) (\left\vert{C}\right\vert) -
\nabla(2-U_j^c-X_i^c) \dots \forall i,j \in N, \forall c\\
\intertext{If the $\nabla$ related term is neglected, then the false-negative cost is computed in terms of the number of sets that will be needlessly explored. The $\nabla$ related term voids the constraint if $j$ is not the lead element or if element $i$ is not really present. Thus, the constraint addresses the false-negative case when group $c$ is led by element $j$ and the user discards $c$ to search for $i$ because of a low value for $E^{ij}$. This means a high value for $(1-E^{ij})$. Next, we examine handling of the penalty related to the tab locations where certain commands are normally expected.} 
&\Omega_i^\tau \ge \lambda_4 (1-Y_i^{\tau*}) \dots \forall  \dots \forall i \in
N, \forall c\\
\intertext{Here, $\tau*$ refers to the preferred tab as specified by $L_i$. If the command is not on its preferred tab, a penalty of $\lambda_4$ is incurred.}\notag
\end{align}

\subsection{Handling of loners}
``Loner'' commands have poor association with other commands but
may have a high frequency of usage. To avoid disturbing the cohesion of other
(well-associated) groups, we strive to put all such loners in a
separate group of their own. 
However, the loner group itself becomes contentious when the association for
a specific command is not at either extreme -- that is, when the command is not
associated strongly with other commands but does not actually have an average association low enough to denote it as a loner. Putting such commands in the loner group makes this
loner group too large. 

We introduce a new hypothetical (invisible) command $\kappa$ to the set $N$. This command
will not be placed in the actual menu; it is only introduced temporarily to serve as a focal point
for association of all loner commands. The new command has
the lowest non-zero usage frequency and no location preference, but it will still be constrained to be the lead element for its group, specifying $\sum_c U_\kappa^c = 1$.  The association of this command with all other commands is computed as follows:
\begin{enumerate}
  \item For any command $i \in N$, compute the sum of the association scores for $i$
  with all other commands $j \in N$, and designate this sum as $\Sigma_i$
  \item Find the largest value of $\Sigma_i$ among all $i \in N$, and designate this
  maximum as $\nabla$
  \item For any command $i \in N$, compute the value $\nabla - \Sigma_i$, to be
  designated as loner factor $\mathbb{W}_i$
  \item \label{TheWeightageStep} The association of command $i$ with $\kappa$ is
  calculated as $\mathbb{W}_i$ / the square root of $N$
  \item If any command $i$ has greater than average association with any one command
  in $N$, the association of command $i$ with $\kappa$ is marked as zero 
\end{enumerate}
After this, the optimization problem is solved for the augmented set $N +
\{\kappa\}$. Naturally, the association score for strongly loner elements with
$\kappa$ is quite high. This ensures that $\kappa$ serves as a ``loner magnet,''
attracting all loner elements to a single group. Further, the weight 
used to compute the association score from the loner factor in step
\ref{TheWeightageStep} can be modified to control the size of the loner group. No other constraints are required for the loners. The loner factor automatically handles the balance of populating the relevant group with commands while avoiding the problems of very large or very small groups.

\subsection{Parameter values} \label{Parameter values}
The results from IFT (and even from the minimal representative formulation) are sensitive to the values set for $\lambda$,
as direct weights in the objective function, any gross modifications to these values result in different menu layouts. 
For the results reported on in this paper, 
we have aimed for rough equivalence in settings among the objectives: no single term dominates in the overall objective function. To achieve this equivalence, we still need some rough information about the expected resultant layout of menus. So, we used the following as initial indicative intentions regarding layouts:
\begin{enumerate}
\item We prefer the menu layouts such that the number of commands (rows) in every tab is not varying too much. As an example, we note that the baseline or commercial menu structure in Notepad application has a very strong variation. While the \emph{Edit} tab has $11$ commands, the \emph{View} tab has just one command. We would not prefer such a huge variation in our menus.
\item The maximum number of permitted tabs scales up as a logarithmic function of the number of commands to be placed. 
\item We prefer not to use single-element sets. If some command is so unrelated to all other commands, it is a suitable candidate for being in the loner set. A single element group -- as used in the \emph{View} tab of the baseline version of Notepad application -- will create several problems. Primarily, it hampers the access time of all subsequent commands. Secondly it also increases the total cost of all false negatives for previous commands.
\end{enumerate}
The intentions listed above are not enforced as hard constraints and are not included in the objective functions. Rather, they are used as indicative guidelines while designing the values of parameters $\lambda$. 

Consider the values of $\lambda_f$ and $\lambda_c$ as presented above in Equation \eqref{MRF_Objective} for the minimal representative formulation. For a data instance involving $n$ elements, we first expect that roughly $\log(n)$ tabs are allowed and every tab has $\frac{n}{\log(n)}$ commands on average. Assuming that the overall User-Interface is of width $w$ and height $h$, the term $\sum_{i\in N} \mathbb{F}_i t_i $ in Equation \eqref{MRF_Objective} will have a value of the order of magnitude of roughly $\frac{wnh}{2\log(n)}$. The term $\sum_{i\in N} \sum_{j\in N} \mathbb{A}_{ij} {Z}_{ij}$ will be in the general vicinity of $n^2/2$. For a sufficiently large canvas and for lower value of $n$, value of $\frac{wnh}{2log(n)}$ is substantially larger than $n^2/2$. To ensure that the two terms have comparable impact in the overall objective function, the value of $\lambda_c$ should be around $\frac{2wh}{nlog(n)}$ times that of $\lambda_f$.
A similar approach is used to initially set the lambda values for IFT. 

However, we expect that the designer will fine-tune the results further. It is expected that the final relative values will be ascertained through trial-and-error over a large number of results.

\section{Results} \label{Results}In this section, we assess the quality of the optimized designs and report on performance quantitatively.

\subsection{Task instances} \label{Task instances}
The approach was tested with three realistic design scenarios, selected for their representativeness, size, and diversity as cases. 
Our goal was to use a wide range of often-used applications, where the names and meanings of most commands are clear to typical users. 
Users who are already conversant with the commands will also have some ideas and expectations as to mutual associations, preferred
placements, etc. With these objectives, 
we chose the following applications as test cases:
\begin{enumerate}
  \item The classic \textit{Windows Notepad\texttrademark}
  application is a widely used compact, elementary text editor. The existing design involves 23 commands, distributed across five tabs.
  \item The \textit{Adobe Acrobat Reader\texttrademark}
  application is a commonly used reader for files in PDFformat. Version 11 of this application has a menu system with 46 commands, distributed over five tabs. 
  \item \textit{Mozilla Firefox\texttrademark 3.6} is a well-known browser application. Its menu system has 51 commands, spread across seven tabs.
\end{enumerate}
These instances are represented by means of the following parameters:
(i) names (text strings) and relative frequency values $\mathbb{F}$ for all
 commands, (ii) association score $\mathbb{A}$ for every pair of commands, and (iii)
 (optional) location preference $\mathbb{L}$ for any of the commands.
 
To ensure the practicality of the instances, the authors implemented
a short exercise in which two external designers (students enrolled in an human-computer interaction program) were asked to rate two parameters independently:
(i) how many times in the course of a typical usage session is a specific command required and (ii) how closely are the two given commands related to each other. The 
answers were quantified to yield the frequency and the association
score, respectively. We found the disagreement to be below one percent, so the values from the two raters were accepted as the data instance specification.

In the discussion below, the incumbent menu is the menu that exists in the commercial applications as of the time of writing. This is referred to as the ``baseline design.'' The baseline designs for all data instances are provided in Appendix \ref{Baseline Designs for all Data Instances}.  
The layouts produced via the optimization method proposed in the paper are
denoted as the ``optimized design.'' Optimized designs are covered in Section \ref{Results}.

\subsection{Implementation and Numerical Performance}
The mixed-integer programming (MIP) formulations were coded in the Java\texttrademark\ SE (build 1.8) plarform. These formulations were solved using
IBM\texttrademark \ Cplex\texttrademark\ 12.6.2 solver on an eight-core 64-bit Intel\texttrademark\ i7\texttrademark\ processor running at
2.8~GHz with 16~GB of RAM. The MIP solver was used with Concert\texttrademark\
technology in conjunction with several customized callbacks.

Computation times for the three cases are given in Table \ref{TablecomputationalTime}. These times are reported as averages over multiple computational executions, with different
values for weight functions $\lambda$ and $\Phi$ (for values as computed in Subsection \ref{Parameter values}). 

We note that for extreme values of
$\lambda$ and $\Phi$, much shorter computation times were observed. For example, if $\lambda_f >> \lambda_c$ and $\lambda_f >> \lambda_m$ are set, then the minimal representative formulation  objective does not really consider the association between commands at all. For such an extreme setting, the corresponding problem can be solved within a small fraction of the computational time listed. So, the computational performance strongly depends on the chosen parameters.

The listed performance is for the default versions of the MIP formulations (where the raw model is passed to the solver without making any attempts to improve performance). In practice, there exist several MIP techniques to improve computational performance. Further, the performance improves substantially when we start with a known (existing) solution instead of starting from scratch. In our case, one feasible solution (the existing layout) is always known, so this technique can be easily utilized. However, the discussion of such MIP techniques is beyond the scope of the current paper; our focus is to demonstrate the efficacy of the menu design approach and not the numerical method.

\begin{table}[!htbp]
\small
\centering
\caption{Computation times for optimal solutions from the optimizer, for two distinct evaluative functions}
\scalebox{0.8}{
\begin{tabular}{l|c|c|c}
    \hline
    \hline
    \multirow{2}{*}{Application} &
    \multirow{2}{*}{Number of elements} &
    \multicolumn{2}{c}{Average time to find the optimal
    solution, in minutes} \\ \cline{3-4} 
  &  & Two-fold-objective Approach & Information Foraging Approach \\
    \hline
    Windows Notepad & 23 & 17  & 22 \\
    Adode Acrobat Reader & 46 & 960 & 1201 \\
    Mozilla Firefox & 51 & 1145 & 1633 \\
    \hline
    \end{tabular}}%
  \label{TablecomputationalTime}%
\end{table}%

Although the computational effort excludes the use of these solvers in interactive design tools, they appear satisfactory with regard to the problem considered here, especially for one-shot optimization.
Developers of professional software should be ready to invest a few hours or even two weeks of computational effort to ensure the quality of their menu design.
This can be further speeded up by using dedicated hardware: our computations were done using a commodity laptop.
Secondly, the data shows that using IFT does not significantly impair performance
relative to that with the earlier, two-fold-objective approach.

\subsection{The two-fold-objective approach}
\begin{figure}[!htbp] \centering
\scalebox{0.5}
{\includegraphics{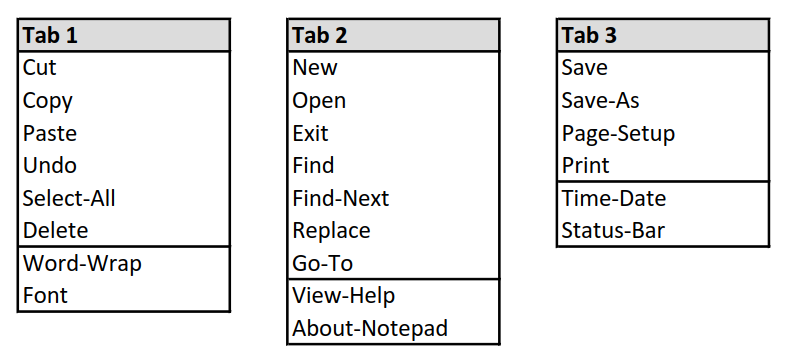}}
\caption{A menu optimized for the Windows Notepad\texttrademark\ text editor with the two-fold-objective minimal representative formulation approach.}
\label{NotepadClassical.PNG}
\end{figure}

\begin{figure}[!htbp] \centering
\scalebox{0.4}
{\includegraphics{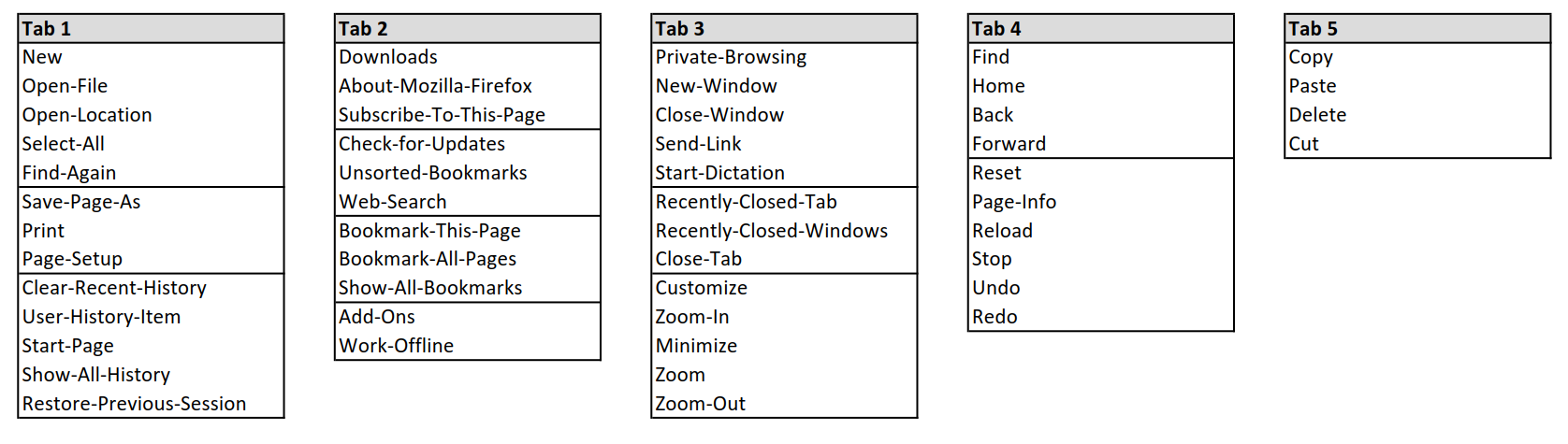}}
\caption{A menu optimized for the Mozilla Firefox\texttrademark\ browser with the two-fold-objective minimal representative formulation approach.}
\label{FirefoxClassical.PNG}
\end{figure}

\begin{figure}[H] \centering
\scalebox{0.4}
{\includegraphics{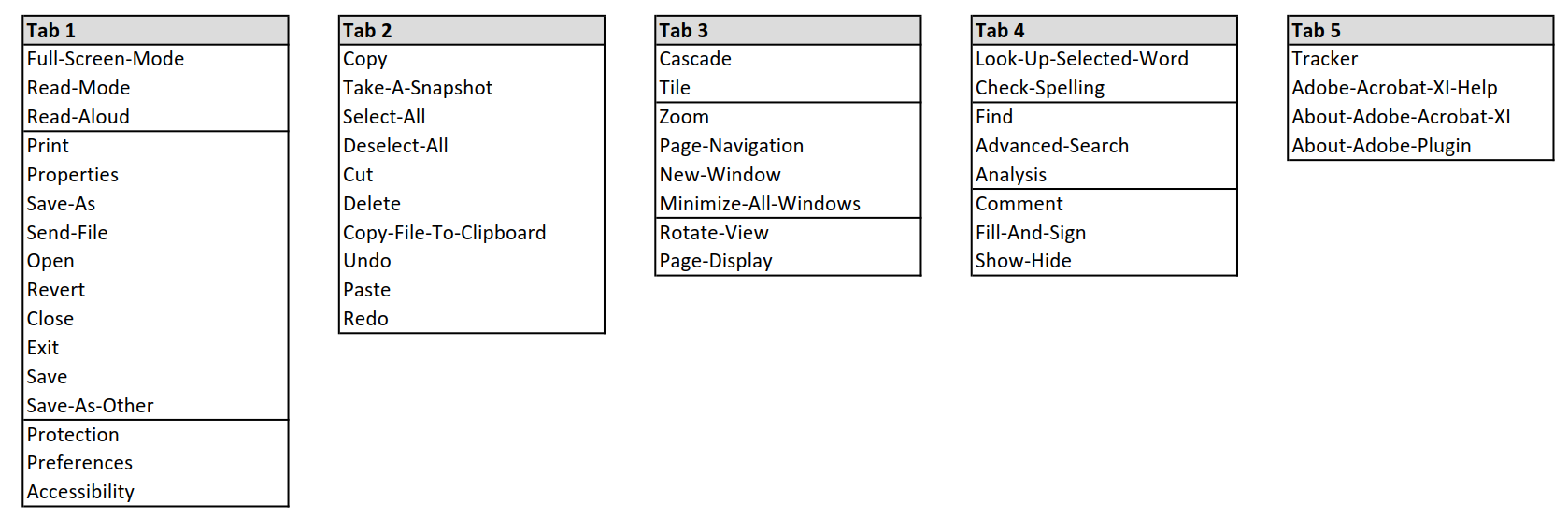}}
\caption{A menu optimized for the Adobe Acrobat\texttrademark\ PDF viewer with the two-fold-objective minimal representative formulation approach.}
\label{AcrobatClassical.PNG}
\end{figure}

We now present designs generated via the algorithms discussed earlier for the data instances specified in Subsection \ref{Task instances}. 
The results for the three data instances with
the two-fold-objective approach are provided in figures \ref{NotepadClassical.PNG},
\ref{FirefoxClassical.PNG}, and \ref{AcrobatClassical.PNG}.

It should be noted that the optimized menu for Notepad has fewer tabs than the baseline (commercial) design. 
This is because the Notepad data instance
involved high scores for association between commands. 
The association values for pairs of
commands for Acrobat and Firefox were comparatively low. The sparse
association led to a wider layout with more tabs and smaller groups. The 
argument extends to the Acrobat menu also: two large groups can be seen in
the Acrobat design, in the first and the second tab. The constituent commands of these groups
have the strongest mutual associations. 
In addition, the association
scores of these commands represent the only case of associations being relatively
large for that data instance. Hence, the two-fold-objective formulation is strongly
driven by the relative association values. 
If we had set the $\lambda$ values in Subsection \ref{Classical Objective Function for MRF} differently, that formulation would not have valued the associations scores so highly. The two-fold-objective formulation is quite sensitive to the $\lambda$ values.
Further trials with the optimizer showed that multiple, quite different
feasible solutions could be found within a narrow range of objective values.
However, the specific relative weightage of the two primary terms in the
objective function led to a situation in which an odd/unexpected placement was linked with
a slightly higher objective value. 
We
conclude that the two-fold-objective nature of the minimal representative formulation  approach (with weighted
performance from Fitts' law and mutual association of commands) led to a more opaque objective function.

Subjectively, it is difficult to justify the placement of a
specific command in the location suggested by the optimal solution.
Some observations can be made nonetheless.
For example, the lead elements of most
groups in the Adobe Acrobat menu are quite esoteric and unrepresentative. 
We note also that the two-fold-objective approach did not, without further information, address preferential
placement of commands on desirable tabs. 
For example, the last/rightmost tab
in the Firefox menu contains the \emph{Cut}, \emph{Copy}, and \emph{Paste} commands and not the
\emph{Help} and \emph{About} commands commonly expected here.  

\subsection{The IFT-based approach to optimized menu designs}

Next, we consider the results obtained with the information foraging approach. The
layouts generated are depicted below, in figures \ref{NotepadResult.PNG},
\ref{FirefoxResult.PNG}, and \ref{AcrobatResult.PNG}.

\begin{figure}[!htbp] \centering
\scalebox{0.5}
{\includegraphics{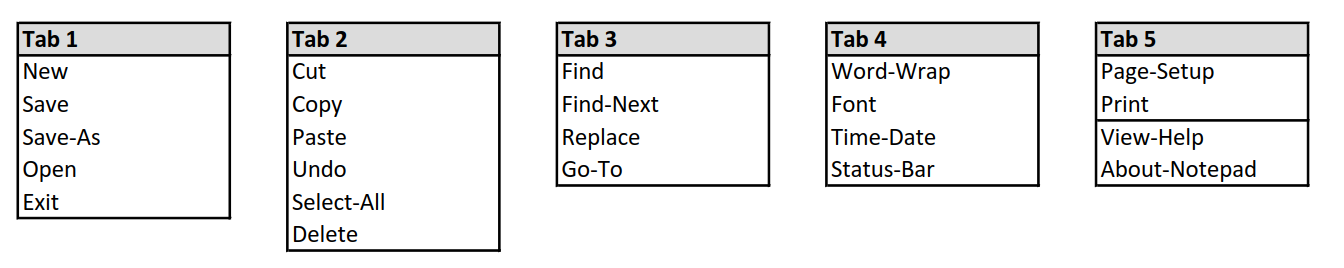}}
\caption{The IFT-based approach for a menu optimized for the Windows Notepad\texttrademark\ text editor.}
\label{NotepadResult.PNG}
\end{figure}

\begin{figure}[!htbp] \centering
\scalebox{0.4}
{\includegraphics{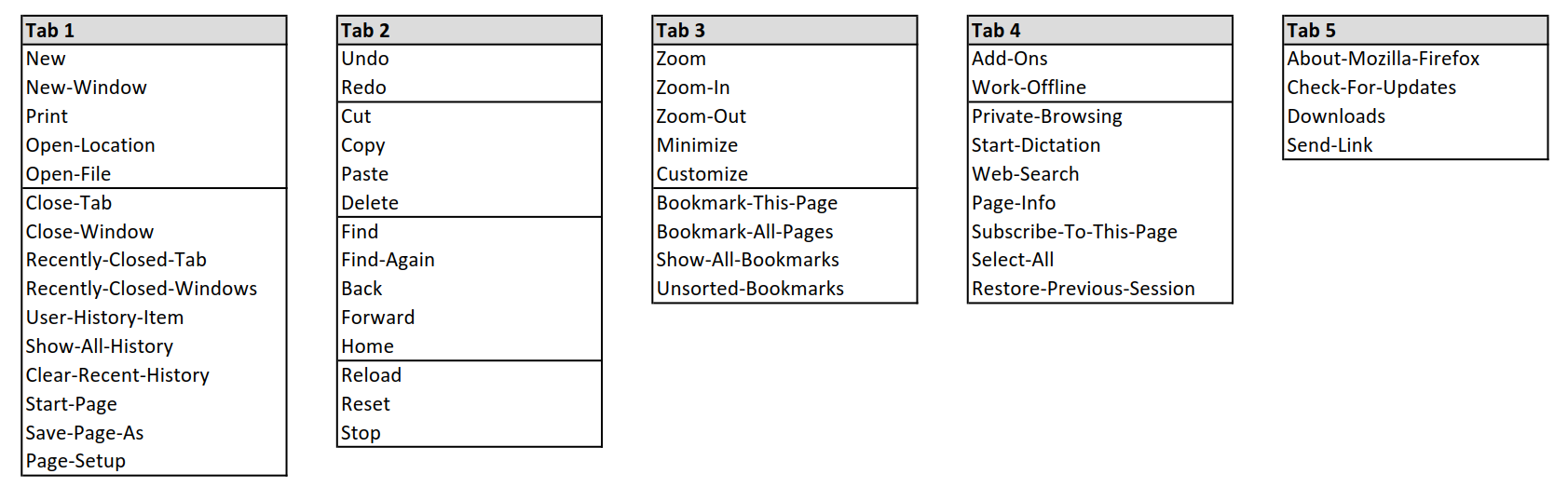}}
\caption{A menu optimized for the Mozilla Firefox\texttrademark\ browser via the IFT-based approach.}
\label{FirefoxResult.PNG}
\end{figure}

\begin{figure}[!htbp] \centering
\scalebox{0.4}
{\includegraphics{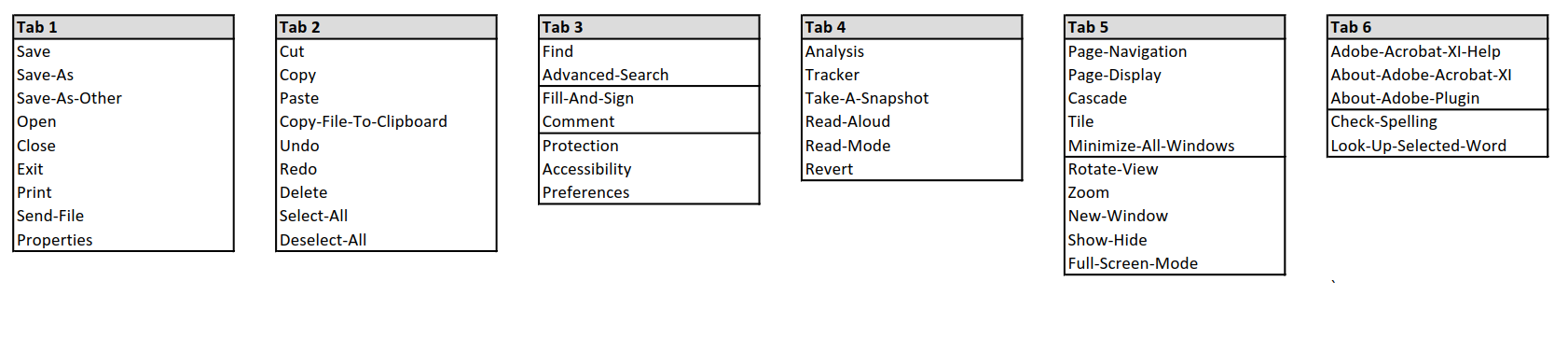}}
\caption{A menu optimized for the Adobe Acrobat\texttrademark\ PDF viewer by means of the IFT-based approach.}
\label{AcrobatResult.PNG}
\end{figure}

A few subjective observations can be made about the results.
Firstly, the results for Notepad and also for Adobe Acrobat show a larger number of tabs. 
However, the tabs appear more balanced than with the previous method.
In particular, there is less deviation among the tabs in the number of commands.
An exception is the first tab in the Firefox menu, which is large relative to the others, 
but the remainder of the menu is well-balanced. 
Also, the groups appear reasonable in light of usage frequency (e.g., \emph{Undo}, \emph{Cut}, and \emph{Copy}), associations between commands (e.g., \emph{Undo} and \emph{Redo}), and expectations for command position (e.g., \emph{About}). 
In addition, groups' first items (lead items) are generally more indicative of the rest of the group here than with the two-fold approach (e.g., ``Full-screen-Mode'' in Figure 4).
Finally, there is good control of loners and also of preferential placement of elements on tabs,
thanks to these being explicitly addressed in the formulation.  
With the next section, we explore whether these observations correlate with empirical results for user performance.

We should note that the IFT-based approach also turned out to be less sensitive than the two-fold one.  
There are relatively few \textit{different} feasible solutions in close proximity to the optimal one.  
The IFT approach is affected less by minor  variations in the relative weights of the terms in the objective function. This should be advantageous in cases wherein the weights cannot be deduced \emph{a priori}.

\section{Empirical Evaluation} \label{User-study}
A controlled laboratory study was carried out in line with established practices in research on menu interaction \citep{bailly2014model}.
We compare average selection times between optimized and non-optimized designs when everything else is kept equal.
In the conditions used, the name of a command (the \emph{target}) is shown on the display and the user is to find and select it as swiftly as possible.
Three optimized designs (the task instances described above) were compared with commercially deployed designs (baselines). 
A non-uniform Zipfian distribution of command selection frequency was used, as in earlier research \citep{liu2017effects}.
We applied the same distributions used in the task instances, which were obtained via data from the external designers (see above).
 To avoid interference effects, 
each user experienced either the optimized or the non-optimized version of a menu, not both.

\subsection{Method}

\textbf{Participants:}
Twenty-four participants were recruited by means of email advertisements and personal networks. Their average age was 29.71 (SD: 3.14). 
Eight of the participants were female. One subject was left-handed. 
All were non-native English-speakers and familiar with computers' mouse and menu systems. At the end of the experimental study, we asked each participant whether he or she had seen these menus before or not, whereupon about 80\% claimed to be familiar with these menus but not to have any idea of whether the locations of the commands had been changed. 
All were compensated with a movie ticket.

\textbf{Experiment design:}
Three applications were studied, as explained in the description of \emph{task instances} above: 
Notepad, Adobe Reader, and Mozilla Firefox. 
Each user used an application either in the optimized or in the baseline condition (again, not in both).
This yielded eight unique combinations (baseline or optimized for Notepad multiplied by baseline or optimized for Adobe Reader multiplied by baseline or optimized for Mozilla Firefox: $2*2*2=8$).  
Participants were assigned to conditions by rotation. 

\textbf{Task and procedure:}
The study started with a brief introduction to the purpose of the study and the tasks to be performed. Demographic data was collected with regard to gender, age, native language, and level of familiarity with menus,  via a questionnaire. After this came the main part of the experiment: The label of every target command (to be searched for) was displayed at the outset,
after which the menu was shown once the participant had pressed a \emph{Start} button. The task now was to find and click the target command as quickly as possible. Selection time was measured as the duration between pressing the \emph{Start} button and clicking the command within the menu. 


For Notepad, Mozilla Firefox, and Adobe Reader, this sequence of steps was completed 40, 80, and 80 times, respectively.
The participants explored one layout at a time before proceeding to the next
application. They had to find several commands within the given candidate layout, and then the next candidate layout was used. The complete procedure took approximately 30 minutes per user.

\textbf{Materials:}
For the 40, 80, and 80 commands (again, presented for Notepad, Mozilla Firefox, and Adobe Reader, respectively),  
the baseline designs were obtained from the latest Microsoft Windows version of the application at the time of the experiment (in January 2018).
We used the Roulette wheel method \citep{goldberg1991comparative} to sample from the frequency distribution of commands in the menu (see the description of the task instances).  
Because the optimizer does not choose tab labels, we used the first command on each tab as the label for that tab. For fair comparison, this was done in both conditions.

\subsection{The apparatus and setup}
The experimental software was implemented in Python with the Tkinter module for the menu system.
Tkinter was used for presenting the menus and for recording selection times, mouse trajectories, and background data. 
The experiment was carried out on a computer running Windows 7, with 8~GB of RAM and a 20-inch LCD display. 
A mouse was used as the pointing device.
The transfer function and other settings of the input device were specified by the experimenter and kept constant across all participants. 

\subsection{Results}
Twenty-seven out of the 960 trials with Notepad, 27 out of the 1,920 with Mozilla
Firefox, and 42 out of the 1,920 with Acrobat Reader  were removed from the
final dataset, for two main reasons: (i) selection of the wrong command
(slip) and (ii) taking excessively long to find the target.  
We found that selection times (STs) were not normally distributed, as
is common in reaction and choice reaction studies, and we used Mann--Whitney
U testing \citep{mann1947test} for the statistical tests.
\begin{table}[!htbp]
\small
\centering
\caption{Statistical results for Notepad, Mozilla Firefox, and Adobe Reader}
\label{my-label}
\begin{tabular}{c|c|c|c|c|c|c|c|c}
\hline
\hline
\multirow{2}{*}{Application} & \multirow{2}{*}{\begin{tabular}[c]{@{}c@{}}Dependent \\Variable\end{tabular}} & \multirow{2}{*}{Type} & \multicolumn{6}{c}{Statistical Values} \\ \cline{4-9} 
 &  &  & Average & SD & Median & U & \multicolumn{1}{c|}{$p$-Value} & \multicolumn{1}{c}{\makecell{Effect\\ Size}} \\ \hline  
\multirow{4}{*}{Notepad} & \multirow{2}{*}{\makecell{Selection \\time}} & {\scriptsize Optimized} & 1.99 & 1.12 &1.63  & \multirow{2}{*}{88984} & \multirow{2}{*}{$<.001$} & \multirow{2}{*}{-0.16} \\ \cline{3-6}
 &  & {\scriptsize Baseline} & 2.38 & 1.58 & 1.93 &  &  &  \\ \cline{2-9} 
 & \multirow{2}{*}{\begin{tabular}[c]{@{}c@{}}Number of\\ tabs selected\end{tabular}} & {\scriptsize Optimized} & 1.43 & 1.15 & 1 & \multirow{2}{*}{59448} & \multirow{2}{*}{$<.001$} & \multirow{2}{*}{-0.40} \\ \cline{3-6}
 &  & {\scriptsize Baseline} & 1.64 & 1.29 & 1 &  &  &  \\ \hline
\multirow{4}{*}{\makecell{Mozilla \\Firefox}} & \multirow{2}{*}{\begin{tabular}[c]{@{}c@{}}Selection\\ time\end{tabular}} & {\scriptsize Optimized} & 3.01 & 2.64 & 2.11 & \multirow{2}{*}{360192.5} & \multirow{2}{*}{$<.001$} & \multirow{2}{*}{-0.17} \\ \cline{3-6}
 &  & {\scriptsize Baseline} & 3.36 & 2.74 & 2.52 &  &  &  \\ \cline{2-9} 
 & \multirow{2}{*}{\begin{tabular}[c]{@{}c@{}}Number of \\tabs selected\end{tabular}} & {\scriptsize Optimized} &1.84  &2.18   &  1 & \multirow{2}{*}{417387.5} & \multirow{2}{*}{$<.001$} & \multirow{2}{*}{-0.08} \\ \cline{3-6}
 &  & {\scriptsize Baseline} & 2.07 & 2.25 & 1  &  &  &  \\ \hline
\multirow{4}{*}{\makecell{Adobe\\ Reader}} & \multirow{2}{*}{\makecell{Selection\\time}} & {\scriptsize Optimized} & 3.37 &2.63&2.42&  \multirow{2}{*}{341381.5} & \multirow{2}{*}{$<.001$} & \multirow{2}{*}{-0.20}\\ \cline{3-6}
 &  & {\scriptsize Baseline} & 4.59 &4.14 & 3.14 &  &  &  \\ \cline{2-9} 
 & \multirow{2}{*}{\begin{tabular}[c]{@{}c@{}}Number of\\ tabs selected\end{tabular}} & {\scriptsize Optimized}  & 2.22 & 2.10 & 1 & \multirow{2}{*}{421775.5} & \multirow{2}{*}{.080} & \multirow{2}{*}{-0.04} \\ \cline{3-6}
 &  & {\scriptsize Baseline} & 2.68& 2.72 &  1 &  &  &  \\ \hline
\end{tabular}
\end{table}
Average ST was 1.99~s (SD: 1.12) for the optimized Notepad and 2.38~s (SD: 1.58) for the baseline Notepad design,
3.01~s (SD: 2.64) for the optimized and 3.36~s (SD: 2.74) for the baseline Firefox design,
and 3.37~s (SD: 2.63) for the optimized and 4.59~s (SD: 4.14) for the baseline Acrobat Reader design. Moreover, the $p$-value for all three applications was less than $0.05$, showing that there is a statistically significant difference between the optimized and baseline STs. In other words, our method was able to decrease STs for these menus.

We also examined the number of tabs selected before finding of each command. 
The average was 1.43 tabs (SD: 1.15) for optimized Notepad and 1.64 (SD: 1.29) for
baseline Notepad. 
The corresponding figures for Firefox were 1.84 (SD: 2.18) for the optimized and 2.07 (SD: 2.25) for the commercial design, and those for Adobe Reader were 2.22 (SD: 2.10) for the optimized and
2.68 (SD: 2.72) for the baseline design. Statistical testing yielded a significant difference in favor of the optimized design in the case of Notepad and of Firefox. The effect was not
significant for Adobe Reader. 
Nevertheless, the average ST for the optimized Adobe Reader was 1.22~s less than the baseline value.

\begin{figure}[!htbp] 
\centering
\begin{tabular}{cc}
   \raisebox{-.5\height}{\includegraphics[width=11cm]{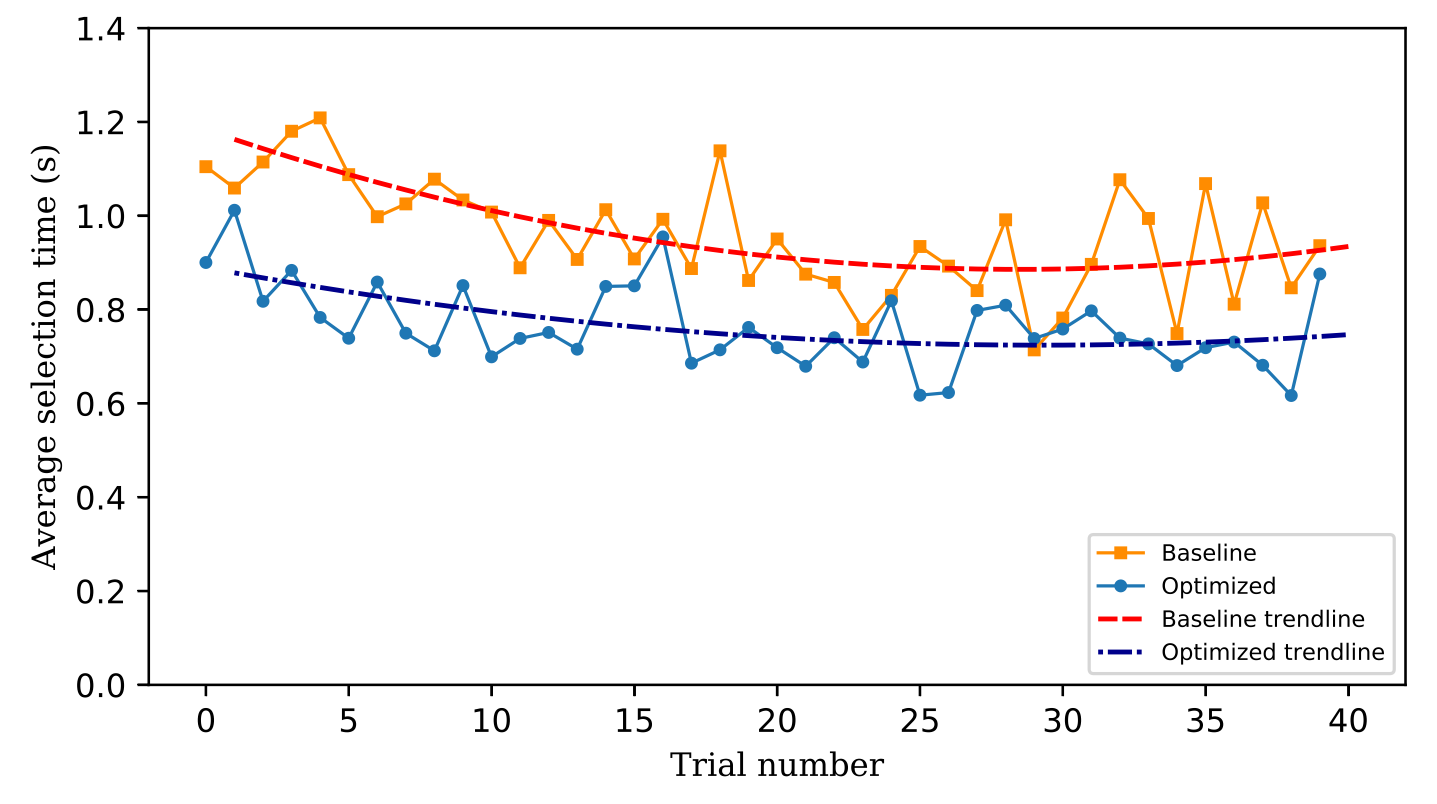}}  &  \rotatebox[origin=c]{-90}{(a) Notepad}\\
  \raisebox{-.5\height}{\includegraphics[width=11cm]{NotepadLC.png}}  &  \rotatebox[origin=c]{-90}{(b) Mozilla} \\
   \raisebox{-.5\height}{\includegraphics[width=11.5cm]{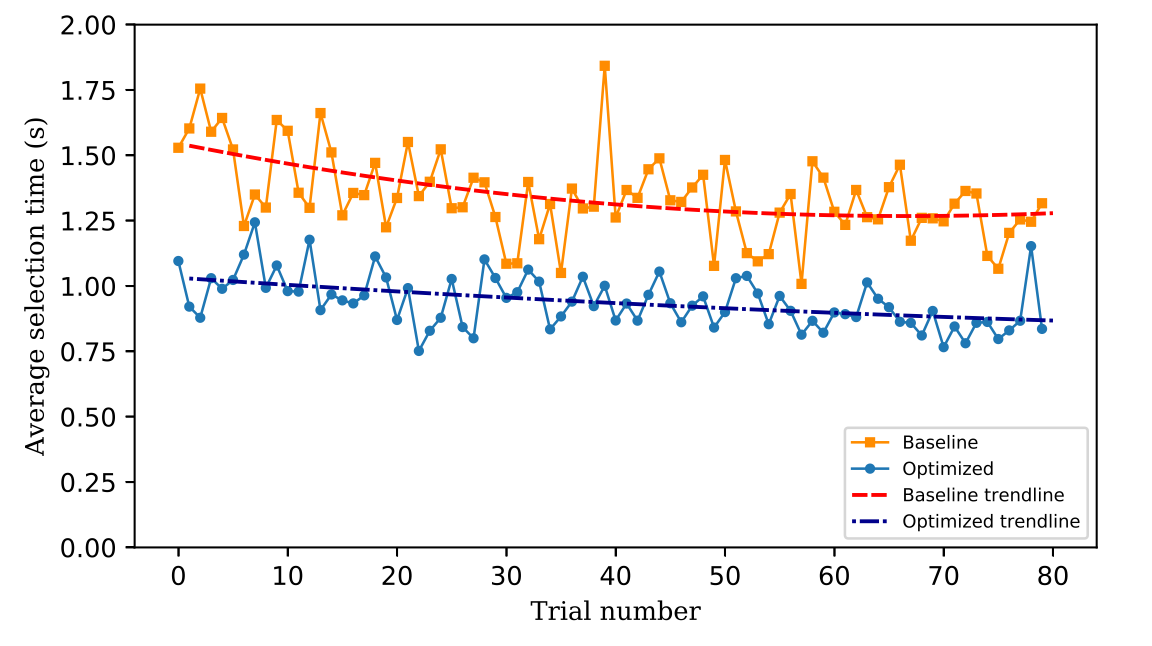}}  & \rotatebox[origin=c]{-90}{(c) Adobe}
\end{tabular}

\caption{Average selection times}
\label{LC}
\end{figure}

The change in selection performance over time is depicted in Figures \ref{LC}. 
All trend-lines in these Figures are default second order polynomial functions. They consistently show a decrease in average selection time for both the baseline and the optimized menu. The optimized one showed better performance at the end of the experiment and in some cases also initially.


\begin{figure}[!htbp] 
\centering \hfill 
\subfigure[Notepad]
{\includegraphics[width=5cm]{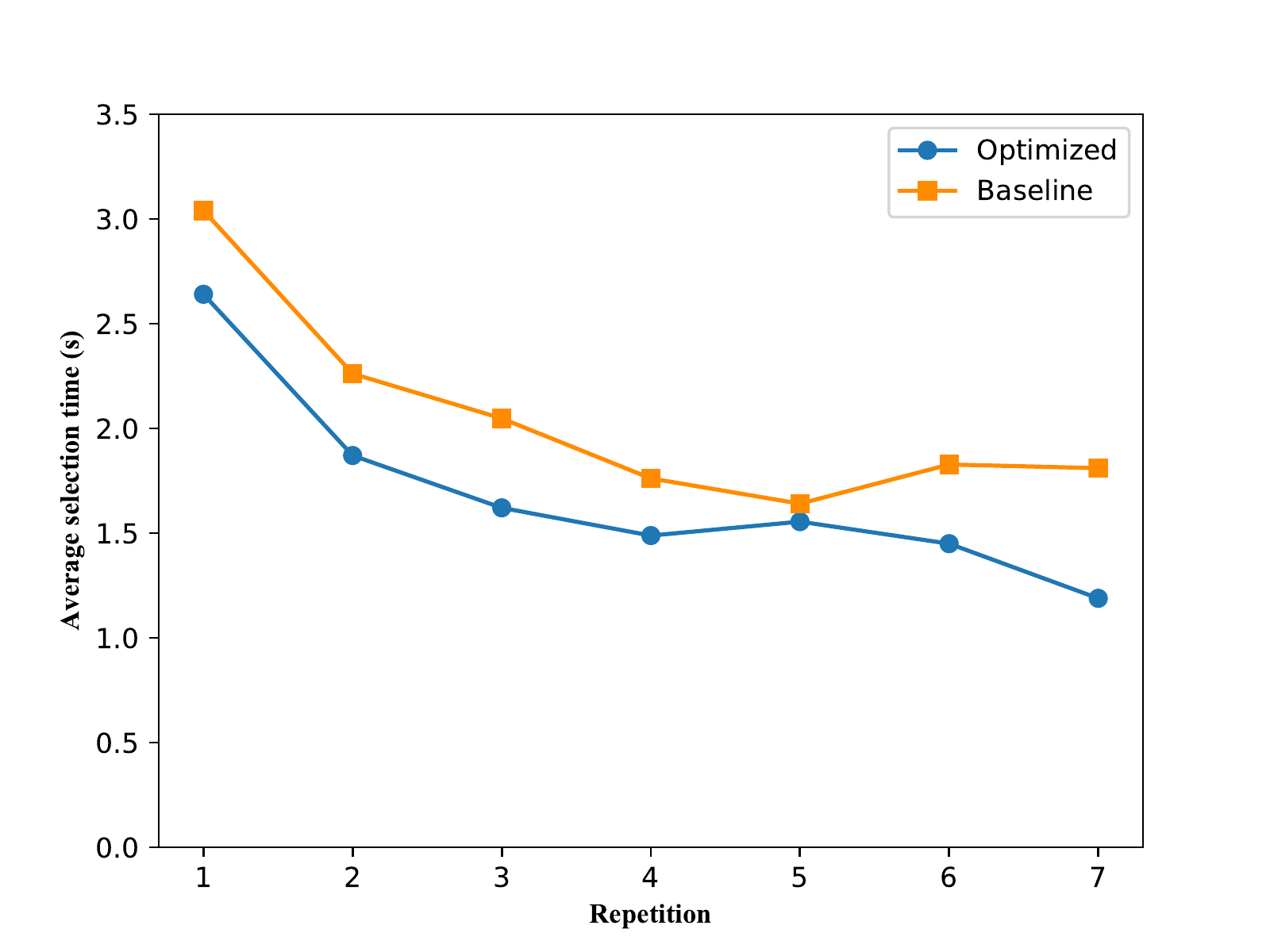}}
\hfill
\subfigure[Mozilla]
{\includegraphics[width=5cm]{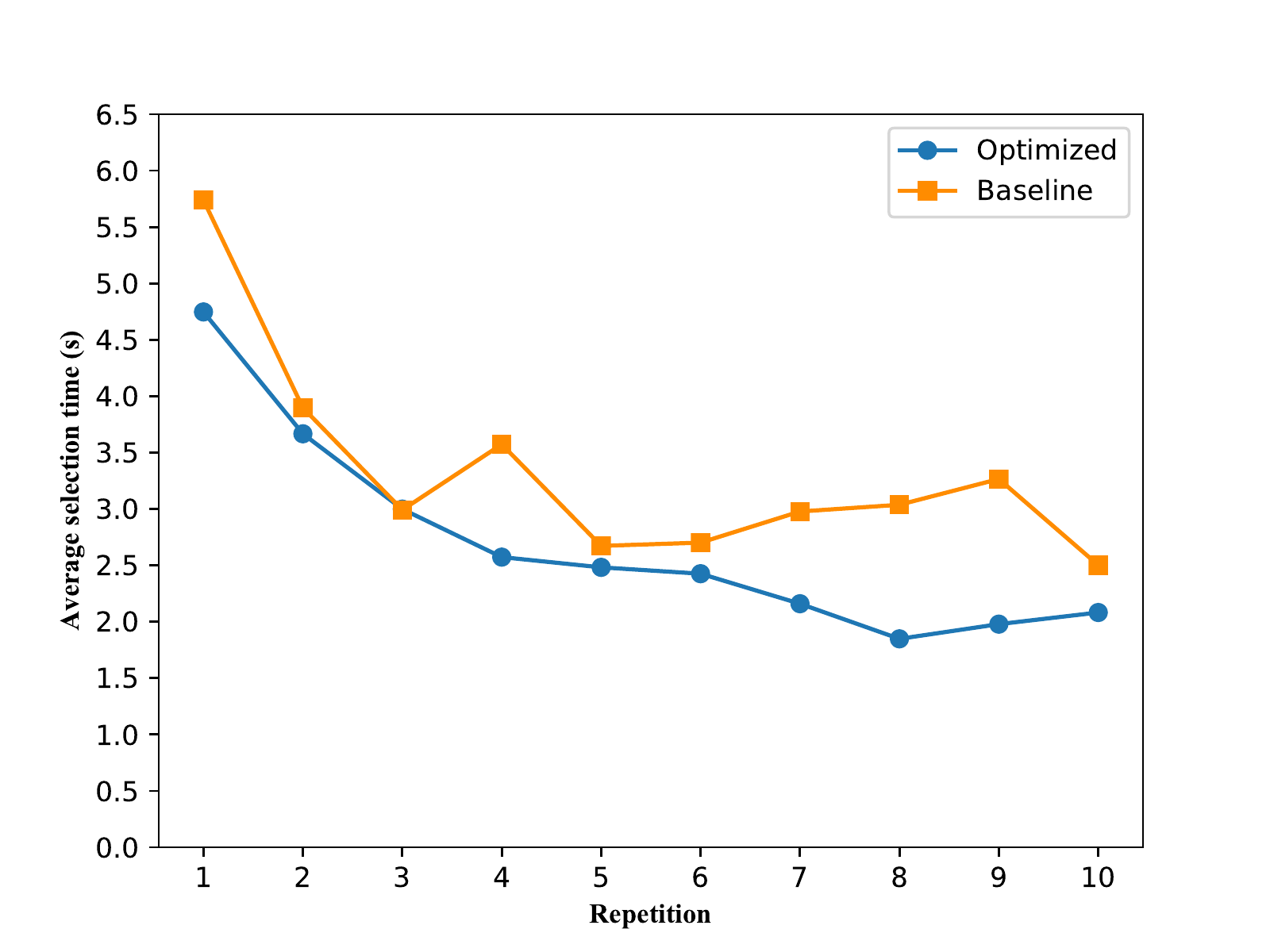}}
\hfill
\subfigure[Adobe]
{\includegraphics[width=5cm]{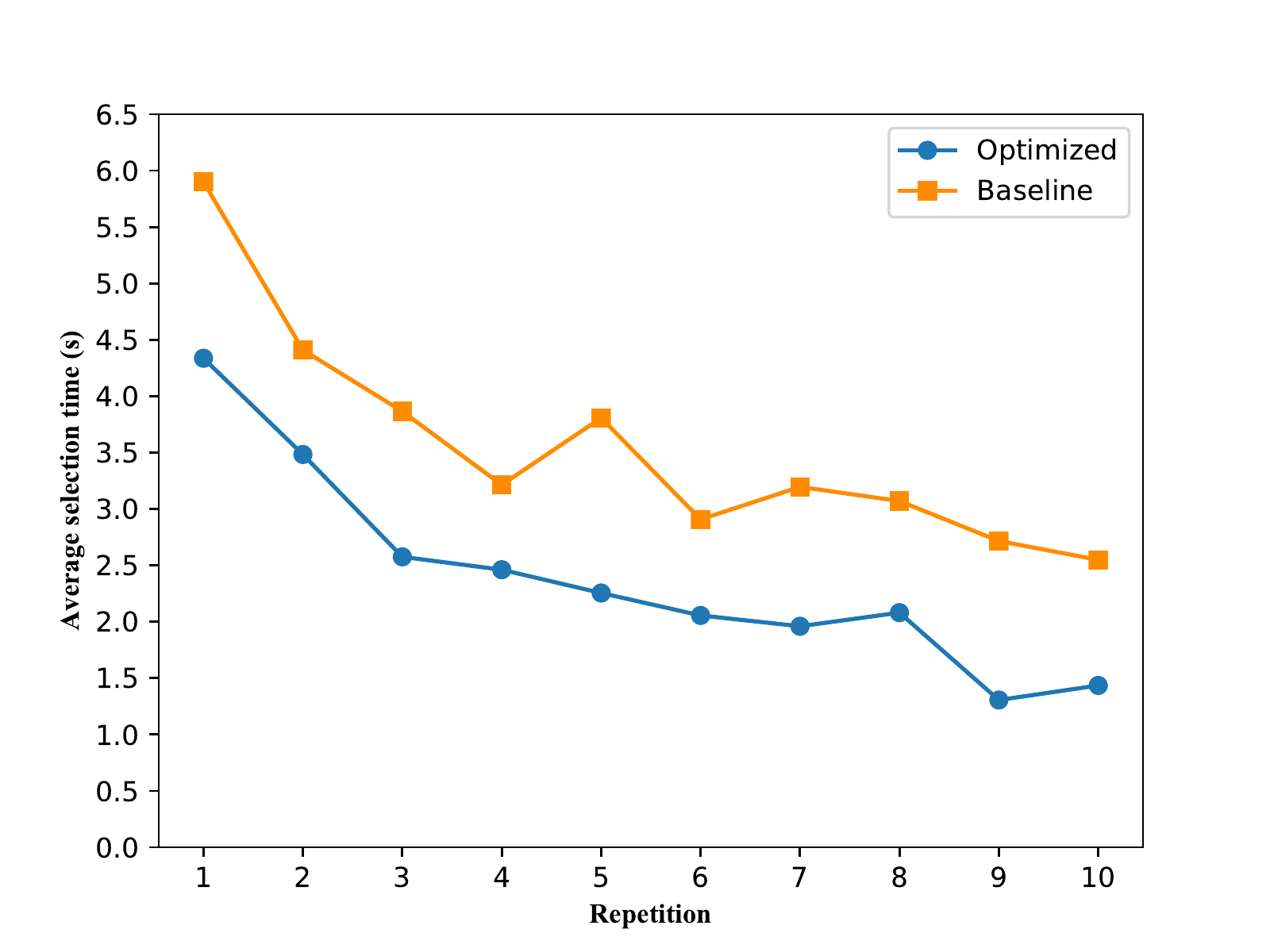}}
\caption{Average selection time per command as a function of number of repetitions}
\label{Graphs_learning}
\end{figure}

To understand learning effect more closely, we report average selection time \emph{per command} in Figure \ref{Graphs_learning}. With increasing repetitions per command, the optimized and baseline menus exhibit similar drop in performance, however the optimized menu shows a more stable trend and an overall lower ST. 

\section{Personalization and Adaptation} 
\label{Personalization and Adaptation} 
The approach is not limited to one-shot computational design.
In this section, we discuss two further applications in personalization and adaptation of a menu system.

\subsection{Personalization} \label{Personalization}
In personalization, 
a menu layout is custom-designed for an individual or a group of users with shared characteristics. 
Some software allows manually customizing menus.
 For example, the Eclipse software system arranges its commands and menus in one way (Perspective) for a "developer" and in a very different way for "tester". 
Our approach makes it possible to do personalization automatically when user data is available.

We illustrate this point using the Notepad application. 
A novice and casual user of Notepad would presumably use the common elementary features such as 'Open, Save, Cut, Copy, Paste' and may also need 'Help' often. On the contrary, an expert would possibly know keyboard shortcuts for most common commands (due to extensive experience and familiarity). 

\begin{figure}[!htbp]
    \centering
\subfigure[Classic (default) Menu layout for Notepad\texttrademark\ application]{\includegraphics[width=.875\textwidth]{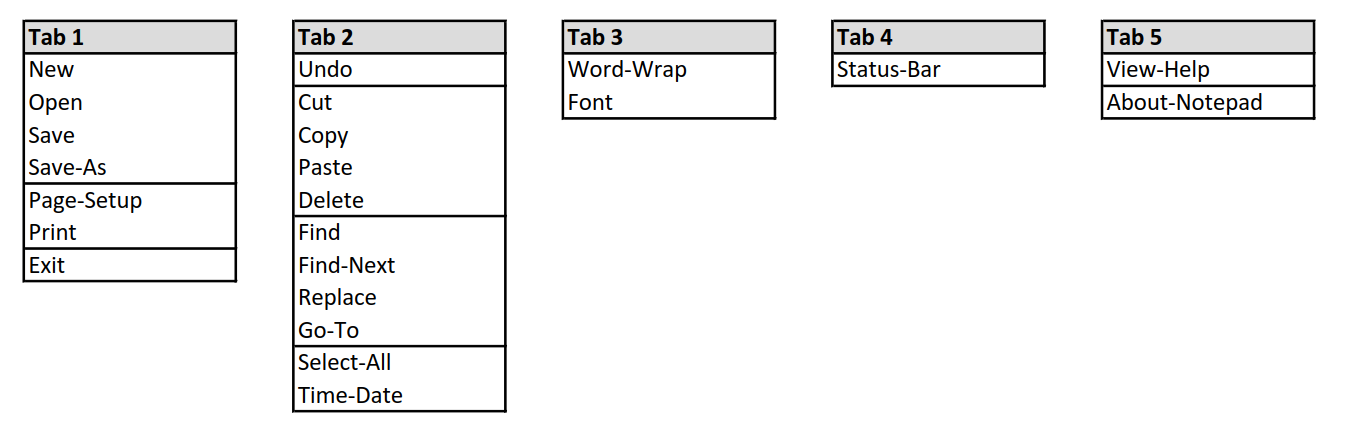}}
\vfill
\subfigure[Menu adapted for a "Novice" user profile.]{\includegraphics[width=.775\textwidth]{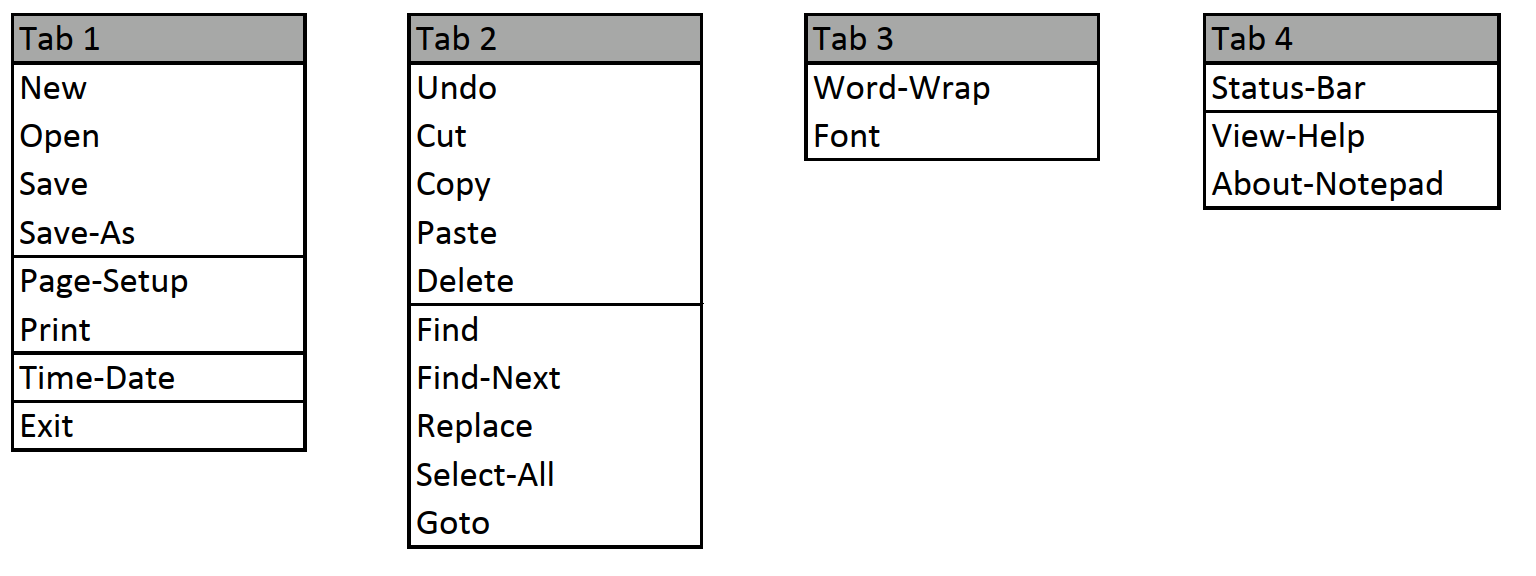}}
\vfill
\subfigure[Menu adapted for a "Expert" user profile. Note the substantial rearrangement where association is sacrificed for performance]{\includegraphics[width=.875\textwidth]{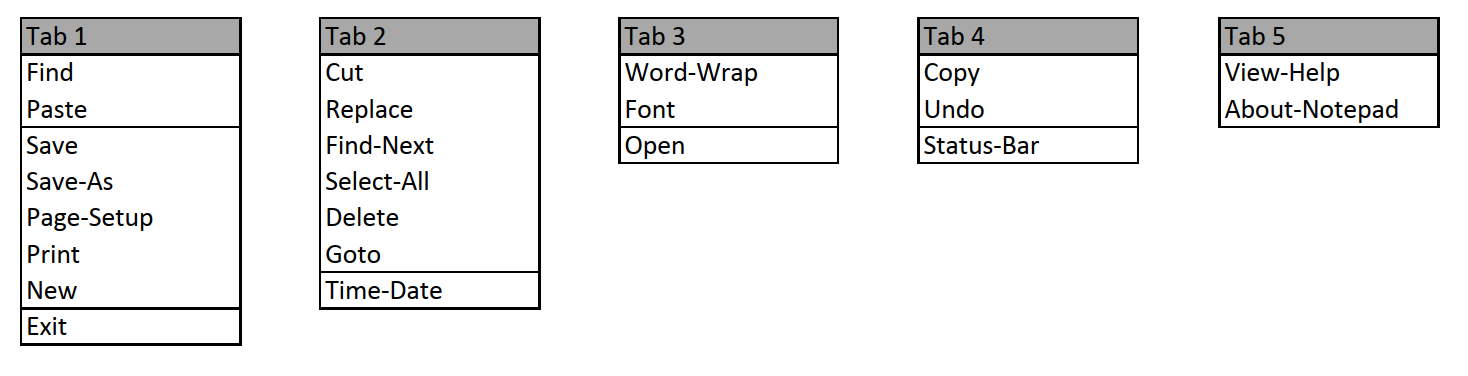}}
\caption{Menu transition from Novice to Expert for Notepad\texttrademark\ application.}
\label{PersonalizedMenu}
\end{figure}

So, we expect that experts would rather only need to use the Menubar to pick rarely used advanced commands such as 'Word-wrap' and 'Font'. This observation leads to two entirely different usage patterns for the two sets of users -- this manifests as two different sets of frequency values for usage of commands. Figure \ref{PersonalizedMenu} shows the menu layouts recommended for a novice and an expert. 

\subsection{Adaptation} \label{Progressive/Gradual Adaptation}

We can also support gradual adaptation of menus. Given a menu and user data, we propose local changes to it in order to improve usability and simultaneously make it learnable. 
Consider that a specific user is very well acquainted with the menu layout of an application. 
If we propose a new layout that is drastically different from the existing one, the user will require substantial effort in 'unlearning' the previous layout and learning a new one. 

\begin{figure}[!htbp]
   \centering
\subfigure[Default menu layout for Notepad\texttrademark\ application]{\includegraphics[width=.875\textwidth]{Notepad.png}}
\vfill
\subfigure[Close-to-original layout. Performance does not improve significantly, but changes are more modest]{\includegraphics[width=.775\textwidth]{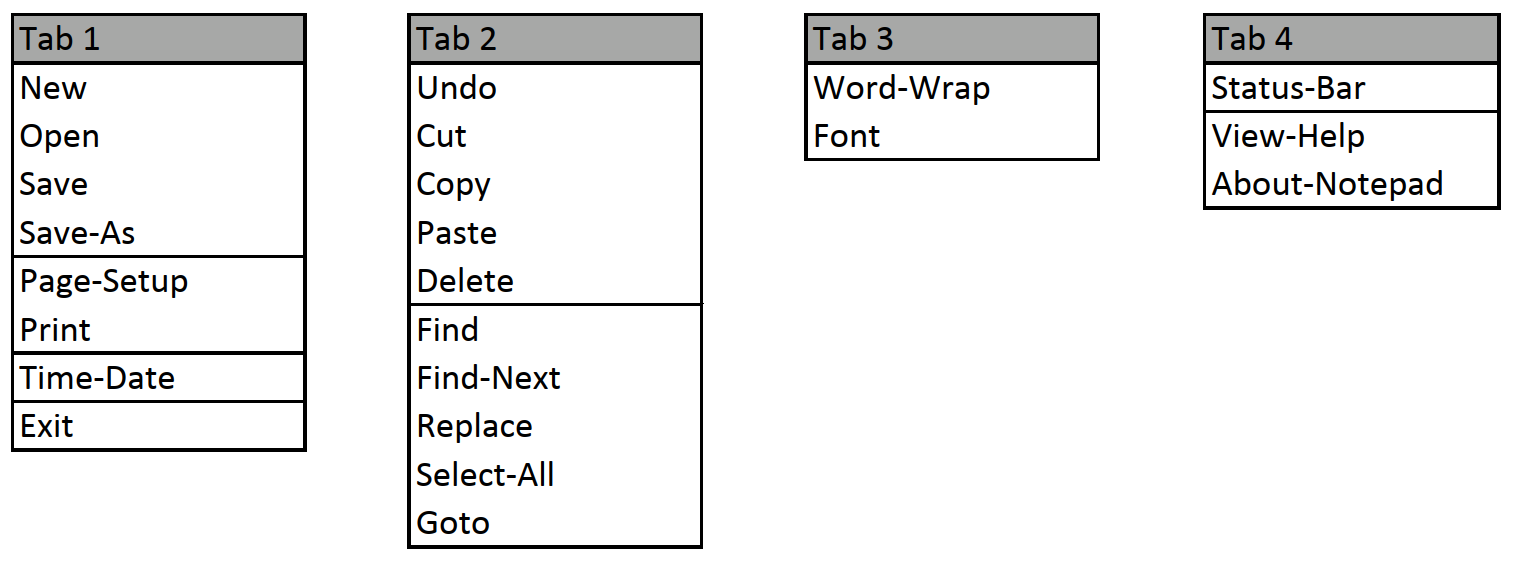}}
\vfill
\subfigure[A farther-from-original layout. Performance improves significantly]{\includegraphics[width=.775\textwidth]{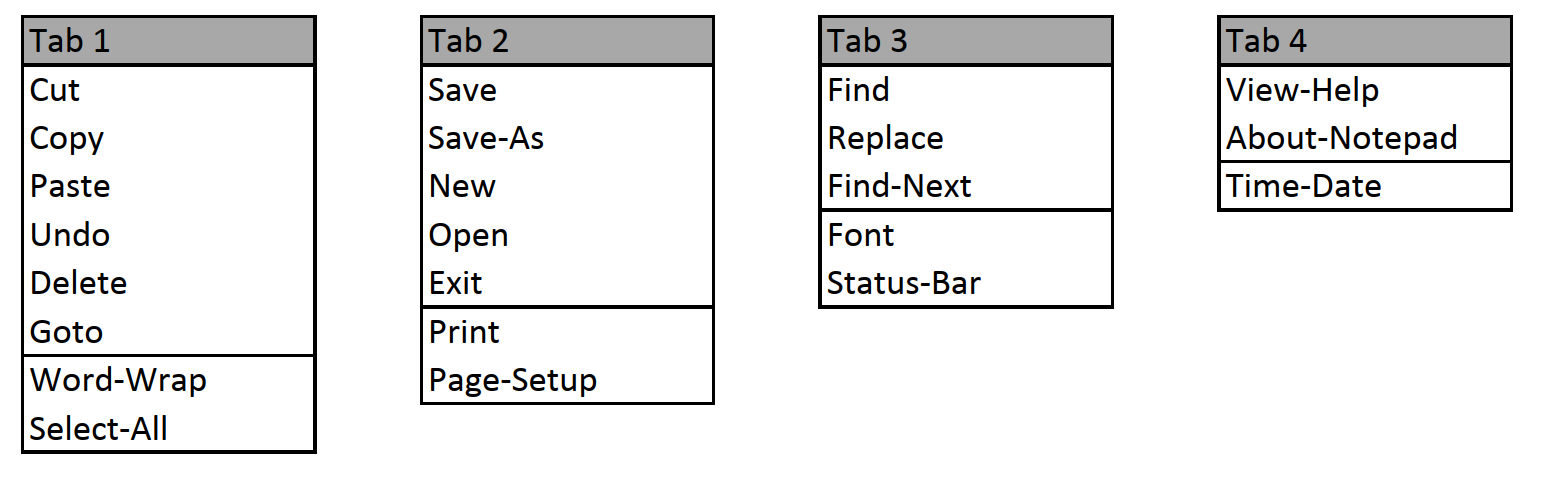}}
\caption{Gradual adaptation for Notepad\texttrademark\ application}
\label{AdaptedMenuNotepad}
\end{figure}

To minimize this retraining effort and still provide a better layout, we propose an adaptation method building on the IFT-based optimizer.
Consider that we quantify the logical difference between two different menu layouts (that involve the same commands). We propose this logical difference as the weighted sum of the tab position change and the row number change of every command. Further, we constrain our computational menu design procedure to search within a specified vicinity of the previous layout. The resulting formulation is based on the following decision variables:
\begin{align}
&\Pi_{i} = \mbox{Change in tab position for command $i$ from original layout to new layout}\notag\\
&\Xi_{i} = \mbox{Change in row position for command $i$ from original layout to new layout}\notag
\end{align}
In conjunction with $\Pi$ and $\Xi$, we also use decision variables from Sections \ref{Minimal Representative Formulation} and \ref{The Information Foraging Approach}. Then we compute the changes in the menu layout by using constraints such as:
\begin{align}
&\Xi_{i} >= \sum_r{rR_i^r} -\Bar{R}_i \ \forall i \in N\\
&\Xi_{i} >= -\sum_r{rR_i^r} +\Bar{R}_i \ \forall i \in N
\end{align}
Here $\Bar{R}_i$ is the row position of command $i$ in the original layout. The objective function includes as additional term to minimize the change from the original formulation, such as: \[
w (\sum_{i\in N}(\Pi_i + \Xi_i)) + (1-w) (\mbox{Performance objective from MRF or IFT})
\]
Here $w$ is a preference term showing proximity to the original layout. This formulation forces the optimizer to find better-performing layouts that are not too much different from the existing layout. Figure \ref{AdaptedMenuNotepad} illustrates this point by showing the Notepad menu gradually changed with increasing distance from the original layout.

\begin{figure}[!htbp] \centering
\scalebox{0.65}
{\includegraphics{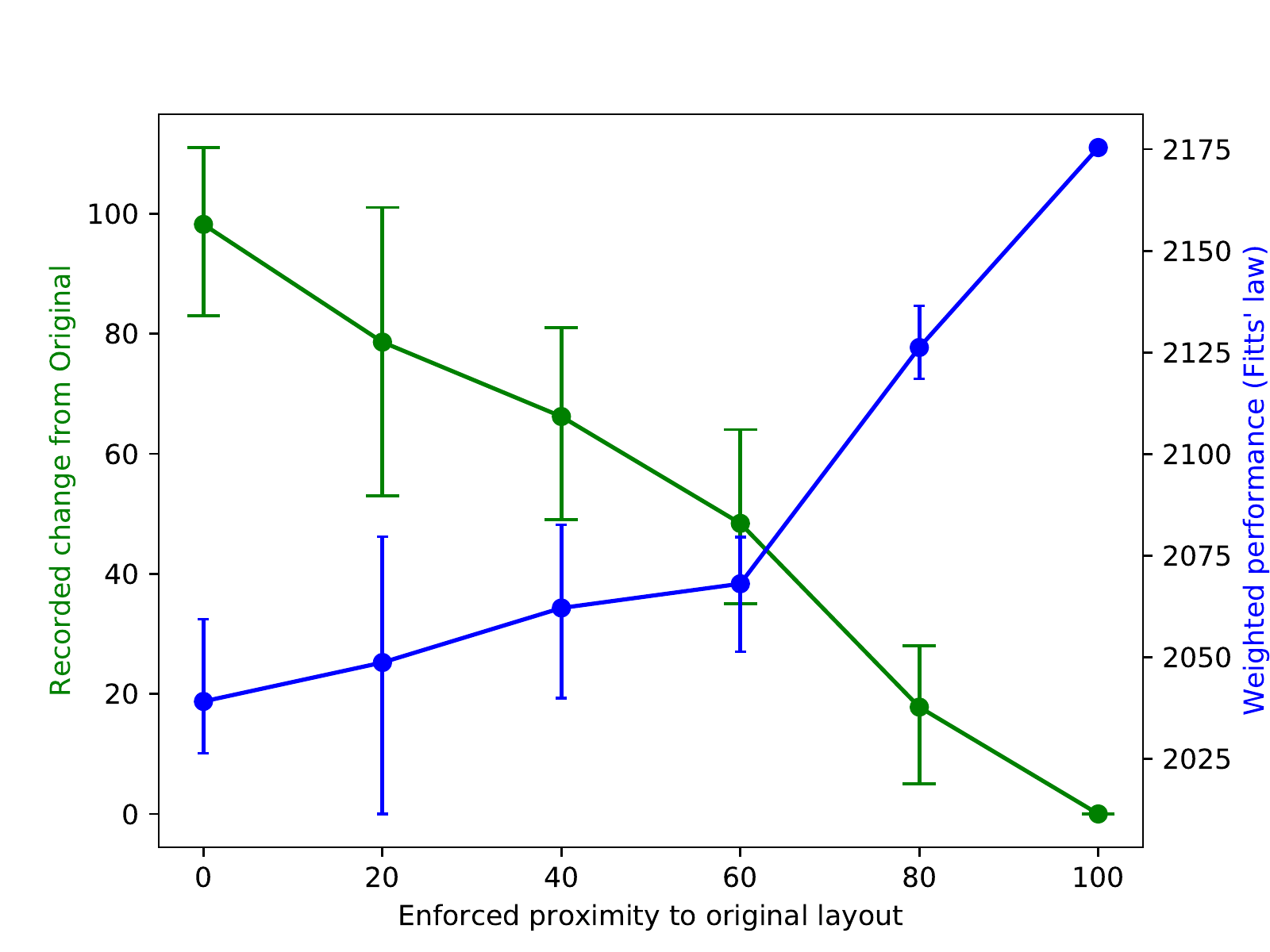}}
\caption{Comparing benefit gained (performance) versus Change from original (loss of familiarity)}
\label{Adaptation.png}
\end{figure}

To appreciate the impact of this gradual adaptation of menu layouts, consider Figure \ref{Adaptation.png}. This shows the gradual adaptation of the menu layout where we balance the proximity to the original layout against the objective of getting better performance. In this Figure, we intend to demonstrate that our approach supports controlling the amount of change made for adaptation of a menu layout. We vary the proximity of an adapted layout to the original layout (OX axis), and see what performance improvement we can achieve with this proximity constraint (OY axis). The blue line shows the improvement in performance value (Fitts' law). The green line shows the loss of familiarity (change from original). The results reported here are for the Notepad menu.

\section{Discussion and Conclusions} \label{Discussion}This paper has contributed to the study of algorithmic methods for computational design of menus.
The design of menu systems strongly impacts the usability and learnability of the computing application. However, currently menu design remains a manual activity almost exclusively; 
there is no well-established or commonly accepted computational technique  to automatically generate or refine menus. 
The absence of an effective mathematical model for a menu hierarchy -- such as the minimal representative formulation (MRF) proposed here -- has made it difficult to test for reliable objectives and to distinguish a good solution from a poor one. 
To the best of our knowledge, no earlier approach provides guarantees regarding the solution quality,
yields a hierarchically organized menu suitable for large command sets,
is not over-determined by previous designs (e.g., on account of a data-driven approach to the objective function),
does not require much input (relative to the frequency of each command and pair-wise association scores),
can be used for one-shot design as well as adaptation,
and is computationally efficient for large menus. 

Our contribution through the MRF is to enable a compact, flexible and purely linear IP formulation for
solving the assignment and set covering problems simultaneously and within
reasonable computation effort, thereby warranting application in regular menu designs. 
The MRF approach provides enough flexibility in defining the objective function to cover the wide variety of factors involved in menu design
(the total number of tabs, the length of individual tabs, the number of groups
on a tab, the length of individual groups, intra-group and intra-tab associativity, frequency of usage of the commands, etc.).
The MRF supports diverse evaluative functions; 
this is valuable for researchers because various hypotheses can be tested with relative ease. 
Specifically, evaluation functions can be expressed that refer not only to the position of any item (like in the assignment-based approach) but also to its membership of a set such as a tab.
We implemented a 'classical' objective from the previous literature and showed that solutions for the resulting mixed-integer programs can be found within reasonable computational effort. The results yielded by the two-fold-objective approach -- while not entirely unreasonable or impractical --  
suffer from a few shortcomings, such as non-intuitive placement of commands. 

Our novel information-foraging-based approach addresses these problems; it addresses users' decisions in zooming in versus skipping menus when searching for a target. The layouts resulting from the IFT-based (information forage theory) approach appear to be more balanced, better organized (especially lead items), and aligned more closely with expectations regarding command placement. Although the original form of IFT involves non-linear models of user behavior, 
we have demonstrated that a simpler, purely linear, MIP-based (mixed-integer programming) approach yields good results within reasonable computational effort. 
Empirical evaluations suggest that computationally produced menus can be on par with commercial designs
as long as the input data (here, frequencies of selection) reflects actual usage.
However, our empirical evaluation assumed that the usage frequencies match those used to run the optimizer.
Future work should explore how sensitive the outcomes are to a mismatch between inputs and actual use.
Moreover, the problem of how to computationally produce labels for formed groups and tabs remains an open problem.

We draw two conclusions from this paper: The immediate conclusion is that
IFT can be used as an evaluative function in computational menu design with good results.
In conjunction with the MRF, this offers a rigorous, coherent yet flexible new framework for computational menu design. 
Our decision variables and the resulting constraints work with standard commercial MIP solvers and do not require any specialized contributions; for example, we do not require any specialized decompositions, relaxations, or column generation techniques (which are often used to enhance MIP performance). 
This opens possibilities for utilizing more complex evaluation functions with relatively lesser effort. 
In future, to make the method available for interactive design tools,
we will explore heuristic variants and relaxations of the IFT-based approach, which may allow interactive-level performance with large task instances. 
Another limitation to be overcome is related to the nature of the task instances. 
Even if filling-in an association matrix constitutes only about an hour's work for a reasonably large menu system, it may not be practical for some use cases (for instance, in agile or rapid development cycles). 
We will explore word embeddings and other machine learning approaches that can automatically discover command-pair associations from data. 
Moreover, the scope of design decisions covered needs more attention. 
Further work should examine other decisions in menu design, such as label selection and shortcut assignment, and expand from tabbed/grouped menus to other types.

In our opinion, the general class of exact numerical optimization techniques holds promise for hierarchically organized user interfaces more generally.
More research is needed to build on this finding.
Many user interfaces are organized as trees or graphs navigated by selecting from proximally available options \cite{pirolli2007information}. 
The MRF offers a natural representation for the key decisions, 
such as which item to assign to which display, in which group.  
Deepening hierarchies can be addressed by recursively adding decisions.
We found also that the sample--discard--explore formulation of IFT captures an essential aspect of a navigating user's decision-making.
This is an improvement over previous work in user interface optimization, which has utilized mainly 
non-hierarchical evaluation functions \cite{oulasvirta2018combinatorial}.
The linear reformulation proposed here is also sufficiently efficient and avoids resorting to meta-heuristic techniques. 
However, more work is needed to address the different modalities of menu access, such as short cuts and context menus, which add redundancy to the optimization problem and require users to learn more complex strategies of menu use.
Moreover, the naming of tabs remains an open problem.
While word embeddings produce good results for pair-wise association scores, finding a descriptive label for a tab may require considering other types of semantics, such as part--whole relationships. 

The code presented in this paper is made available via our project page.

\bibliographystyle{elsarticle-num-names}
\bibliography{Bibliography}

\begin{thebibliography}{46}
\expandafter\ifx\csname natexlab\endcsname\relax\def\natexlab#1{#1}\fi
\providecommand{\url}[1]{\texttt{#1}}
\providecommand{\href}[2]{#2}
\providecommand{\path}[1]{#1}
\providecommand{\DOIprefix}{doi:}
\providecommand{\ArXivprefix}{arXiv:}
\providecommand{\URLprefix}{URL: }
\providecommand{\Pubmedprefix}{pmid:}
\providecommand{\doi}[1]{\href{http://dx.doi.org/#1}{\path{#1}}}
\providecommand{\Pubmed}[1]{\href{pmid:#1}{\path{#1}}}
\providecommand{\bibinfo}[2]{#2}
\ifx\xfnm\relax \def\xfnm[#1]{\unskip,\space#1}\fi
\bibitem[{Bailly et~al.(2017)Bailly, Lecolinet, and Nigay}]{bailly2017visual}
\bibinfo{author}{G.~Bailly}, \bibinfo{author}{E.~Lecolinet},
  \bibinfo{author}{L.~Nigay},
\newblock \bibinfo{title}{Visual menu techniques},
\newblock \bibinfo{journal}{ACM Computing Surveys (CSUR)} \bibinfo{volume}{49}
  (\bibinfo{year}{2017}) \bibinfo{pages}{60}.
\bibitem[{Bailly et~al.(2013)Bailly, Oulasvirta, K{\"o}tzing, and
  Hoppe}]{bailly2013menuoptimizer}
\bibinfo{author}{G.~Bailly}, \bibinfo{author}{A.~Oulasvirta},
  \bibinfo{author}{T.~K{\"o}tzing}, \bibinfo{author}{S.~Hoppe},
\newblock \bibinfo{title}{Menuoptimizer: Interactive optimization of menu
  systems},
\newblock in: \bibinfo{booktitle}{Proceedings of the 26th annual ACM symposium
  on User interface software and technology}, \bibinfo{organization}{ACM},
  \bibinfo{year}{2013}, pp. \bibinfo{pages}{331--342}.
\bibitem[{Ahlstr{\"o}m(2005)}]{ahlstrom2005modeling}
\bibinfo{author}{D.~Ahlstr{\"o}m},
\newblock \bibinfo{title}{Modeling and improving selection in cascading
  pull-down menus using fitts' law, the steering law and force fields},
\newblock in: \bibinfo{booktitle}{Proceedings of the SIGCHI conference on Human
  factors in computing systems}, \bibinfo{organization}{ACM},
  \bibinfo{year}{2005}, pp. \bibinfo{pages}{61--70}.
\bibitem[{Bailly et~al.(2014)Bailly, Oulasvirta, Brumby, and
  Howes}]{bailly2014model}
\bibinfo{author}{G.~Bailly}, \bibinfo{author}{A.~Oulasvirta},
  \bibinfo{author}{D.~P. Brumby}, \bibinfo{author}{A.~Howes},
\newblock \bibinfo{title}{Model of visual search and selection time in linear
  menus},
\newblock in: \bibinfo{booktitle}{Proceedings of the SIGCHI Conference on Human
  Factors in Computing Systems}, \bibinfo{organization}{ACM},
  \bibinfo{year}{2014}, pp. \bibinfo{pages}{3865--3874}.
\bibitem[{Brumby and Howes(2004)}]{brumby2004good}
\bibinfo{author}{D.~P. Brumby}, \bibinfo{author}{A.~Howes},
\newblock \bibinfo{title}{Good enough but i'll just check: Web-page search as
  attentional refocusing},
\newblock in: \bibinfo{booktitle}{ICCM}, \bibinfo{year}{2004}, pp.
  \bibinfo{pages}{46--51}.
\bibitem[{Byrne et~al.(1999)Byrne, Anderson, Douglass, and
  Matessa}]{byrne1999eye}
\bibinfo{author}{M.~D. Byrne}, \bibinfo{author}{J.~R. Anderson},
  \bibinfo{author}{S.~Douglass}, \bibinfo{author}{M.~Matessa},
\newblock \bibinfo{title}{Eye tracking the visual search of click-down menus},
\newblock in: \bibinfo{booktitle}{Proceedings of the SIGCHI conference on Human
  Factors in Computing Systems}, \bibinfo{organization}{ACM},
  \bibinfo{year}{1999}, pp. \bibinfo{pages}{402--409}.
\bibitem[{McDonald et~al.(1983)McDonald, Stone, and
  Liebelt}]{mcdonald1983searching}
\bibinfo{author}{J.~E. McDonald}, \bibinfo{author}{J.~D. Stone},
  \bibinfo{author}{L.~S. Liebelt},
\newblock \bibinfo{title}{Searching for items in menus: The effects of
  organization and type of target},
\newblock in: \bibinfo{booktitle}{Proceedings of the Human Factors Society
  Annual Meeting}, volume~\bibinfo{volume}{27}, \bibinfo{organization}{SAGE
  Publications Sage CA: Los Angeles, CA}, \bibinfo{year}{1983}, pp.
  \bibinfo{pages}{834--837}.
\bibitem[{Mehlenbacher et~al.(1989)Mehlenbacher, Duffy, and
  Palmer}]{mehlenbacher1989finding}
\bibinfo{author}{B.~Mehlenbacher}, \bibinfo{author}{T.~M. Duffy},
  \bibinfo{author}{J.~Palmer},
\newblock \bibinfo{title}{Finding information on a menu: linking menu
  organization to the user's goals},
\newblock \bibinfo{journal}{Human-Computer Interaction} \bibinfo{volume}{4}
  (\bibinfo{year}{1989}) \bibinfo{pages}{231--251}.
\bibitem[{Norman(1991)}]{norman1991psychology}
\bibinfo{author}{K.~L. Norman}, \bibinfo{title}{The psychology of menu
  selection: Designing cognitive control at the human/computer interface},
  \bibinfo{publisher}{Intellect Books}, \bibinfo{year}{1991}.
\bibitem[{Paap and Cooke(1997)}]{Paap1997533}
\bibinfo{author}{K.~R. Paap}, \bibinfo{author}{N.~J. Cooke},
\newblock \bibinfo{title}{Chapter 24 - design of menus},
\newblock in: \bibinfo{editor}{M.~G. Helander}, \bibinfo{editor}{T.~K.
  Landauer}, \bibinfo{editor}{P.~V. Prabhu} (Eds.),
  \bibinfo{booktitle}{Handbook of Human-Computer Interaction (Second Edition)},
  \bibinfo{edition}{second edition} ed., \bibinfo{publisher}{North-Holland},
  \bibinfo{address}{Amsterdam}, \bibinfo{year}{1997}, pp. \bibinfo{pages}{533
  -- 572}.
\bibitem[{Sears and Shneiderman(1994)}]{sears1994split}
\bibinfo{author}{A.~Sears}, \bibinfo{author}{B.~Shneiderman},
\newblock \bibinfo{title}{Split menus: effectively using selection frequency to
  organize menus},
\newblock \bibinfo{journal}{ACM Transactions on Computer-Human Interaction
  (TOCHI)} \bibinfo{volume}{1} (\bibinfo{year}{1994}) \bibinfo{pages}{27--51}.
\bibitem[{Fu and Pirolli(2007)}]{fu2007snif}
\bibinfo{author}{W.-T. Fu}, \bibinfo{author}{P.~Pirolli},
\newblock \bibinfo{title}{Snif-act: A cognitive model of user navigation on the
  world wide web},
\newblock \bibinfo{journal}{Human-Computer Interaction} \bibinfo{volume}{22}
  (\bibinfo{year}{2007}) \bibinfo{pages}{355--412}.
\bibitem[{Pirolli(2007)}]{pirolli2007information}
\bibinfo{author}{P.~Pirolli}, \bibinfo{title}{Information foraging theory:
  Adaptive interaction with information}, \bibinfo{publisher}{Oxford University
  Press}, \bibinfo{year}{2007}.
\bibitem[{Card et~al.(1980)Card, Moran, and Newell}]{card1980keystroke}
\bibinfo{author}{S.~K. Card}, \bibinfo{author}{T.~P. Moran},
  \bibinfo{author}{A.~Newell},
\newblock \bibinfo{title}{The keystroke-level model for user performance time
  with interactive systems},
\newblock \bibinfo{journal}{Communications of the ACM} \bibinfo{volume}{23}
  (\bibinfo{year}{1980}) \bibinfo{pages}{396--410}.
\bibitem[{Pirolli and Fu(2003)}]{10.1007/3-540-44963-9_8}
\bibinfo{author}{P.~Pirolli}, \bibinfo{author}{W.-T. Fu},
\newblock \bibinfo{title}{Snif-act: A model of information foraging on the
  world wide web},
\newblock in: \bibinfo{editor}{P.~Brusilovsky}, \bibinfo{editor}{A.~Corbett},
  \bibinfo{editor}{F.~de~Rosis} (Eds.), \bibinfo{booktitle}{User Modeling
  2003}, \bibinfo{publisher}{Springer Berlin Heidelberg},
  \bibinfo{address}{Berlin, Heidelberg}, \bibinfo{year}{2003}, pp.
  \bibinfo{pages}{45--54}.
\bibitem[{Oulasvirta et~al.(2020)Oulasvirta, Dayama, Shiripour, John, and
  Karrenbauer}]{oulasvirta2020combinatorial}
\bibinfo{author}{A.~Oulasvirta}, \bibinfo{author}{N.~R. Dayama},
  \bibinfo{author}{M.~Shiripour}, \bibinfo{author}{M.~John},
  \bibinfo{author}{A.~Karrenbauer},
\newblock \bibinfo{title}{Combinatorial optimization of graphical user
  interface designs},
\newblock \bibinfo{journal}{Proceedings of the IEEE} \bibinfo{volume}{108}
  (\bibinfo{year}{2020}) \bibinfo{pages}{434--464}.
\bibitem[{Cockburn et~al.(2007)Cockburn, Gutwin, and
  Greenberg}]{cockburn2007predictive}
\bibinfo{author}{A.~Cockburn}, \bibinfo{author}{C.~Gutwin},
  \bibinfo{author}{S.~Greenberg},
\newblock \bibinfo{title}{A predictive model of menu performance},
\newblock in: \bibinfo{booktitle}{Proceedings of the SIGCHI conference on Human
  factors in computing systems}, \bibinfo{organization}{ACM},
  \bibinfo{year}{2007}, pp. \bibinfo{pages}{627--636}.
\bibitem[{Goubko and Danilenko(2010)}]{goubko2010automated}
\bibinfo{author}{M.~V. Goubko}, \bibinfo{author}{A.~I. Danilenko},
\newblock \bibinfo{title}{An automated routine for menu structure
  optimization},
\newblock in: \bibinfo{booktitle}{Proceedings of the 2nd ACM SIGCHI symposium
  on Engineering interactive computing systems}, \bibinfo{organization}{ACM},
  \bibinfo{year}{2010}, pp. \bibinfo{pages}{67--76}.
\bibitem[{Danilenko and Goubko(2013)}]{danilenko2013semantic}
\bibinfo{author}{A.~Danilenko}, \bibinfo{author}{M.~Goubko},
\newblock \bibinfo{title}{Semantic-aware optimization of user interface menus},
\newblock \bibinfo{journal}{Automation and Remote Control} \bibinfo{volume}{74}
  (\bibinfo{year}{2013}) \bibinfo{pages}{1399--1411}.
\bibitem[{Chen et~al.(2015)Chen, Bailly, Brumby, Oulasvirta, and
  Howes}]{chen2015emergence}
\bibinfo{author}{X.~Chen}, \bibinfo{author}{G.~Bailly}, \bibinfo{author}{D.~P.
  Brumby}, \bibinfo{author}{A.~Oulasvirta}, \bibinfo{author}{A.~Howes},
\newblock \bibinfo{title}{The emergence of interactive behavior: A model of
  rational menu search},
\newblock in: \bibinfo{booktitle}{Proceedings of the 33rd Annual ACM Conference
  on Human Factors in Computing Systems}, \bibinfo{organization}{ACM},
  \bibinfo{year}{2015}, pp. \bibinfo{pages}{4217--4226}.
\bibitem[{Oulasvirta and Karrenbauer(2018)}]{oulasvirta2018combinatorial}
\bibinfo{author}{A.~Oulasvirta}, \bibinfo{author}{A.~Karrenbauer},
\newblock \bibinfo{title}{Combinatorial optimization for user interface
  design},
\newblock \bibinfo{journal}{Computational Interaction}  (\bibinfo{year}{2018})
  \bibinfo{pages}{97--121}.
\bibitem[{Anderson et~al.(1997)Anderson, Matessa, and
  Lebiere}]{anderson1997act}
\bibinfo{author}{J.~R. Anderson}, \bibinfo{author}{M.~Matessa},
  \bibinfo{author}{C.~Lebiere},
\newblock \bibinfo{title}{Act-r: A theory of higher level cognition and its
  relation to visual attention},
\newblock \bibinfo{journal}{Human-Computer Interaction} \bibinfo{volume}{12}
  (\bibinfo{year}{1997}) \bibinfo{pages}{439--462}.
\bibitem[{Nilsen(1996)}]{nilsen1996perceptual}
\bibinfo{author}{E.~L. Nilsen}, \bibinfo{title}{Perceptual-motor control in
  human-computer interaction.}, \bibinfo{type}{Technical Report}, MICHIGAN UNIV
  ANN ARBOR DIV OF RESEARCH DEVELOPMENT AND ADMINISTRATION,
  \bibinfo{year}{1996}.
\bibitem[{Hornof and Kieras(1997)}]{hornof1997cognitive}
\bibinfo{author}{A.~J. Hornof}, \bibinfo{author}{D.~E. Kieras},
\newblock \bibinfo{title}{Cognitive modeling reveals menu search in both random
  and systematic},
\newblock in: \bibinfo{booktitle}{Proceedings of the ACM SIGCHI Conference on
  Human factors in computing systems}, \bibinfo{organization}{ACM},
  \bibinfo{year}{1997}, pp. \bibinfo{pages}{107--114}.
\bibitem[{Byrne(2001)}]{byrne2001act}
\bibinfo{author}{M.~D. Byrne},
\newblock \bibinfo{title}{Act-r/pm and menu selection: Applying a cognitive
  architecture to hci},
\newblock \bibinfo{journal}{International Journal of Human-Computer Studies}
  \bibinfo{volume}{55} (\bibinfo{year}{2001}) \bibinfo{pages}{41--84}.
\bibitem[{Cockburn and Gutwin(2009)}]{cockburn2009predictive}
\bibinfo{author}{A.~Cockburn}, \bibinfo{author}{C.~Gutwin},
\newblock \bibinfo{title}{A predictive model of human performance with
  scrolling and hierarchical lists},
\newblock \bibinfo{journal}{Human--Computer Interaction} \bibinfo{volume}{24}
  (\bibinfo{year}{2009}) \bibinfo{pages}{273--314}.
\bibitem[{Lee and MacGregor(1985)}]{lee1985minimizing}
\bibinfo{author}{E.~Lee}, \bibinfo{author}{J.~MacGregor},
\newblock \bibinfo{title}{Minimizing user search time in menu retrieval
  systems},
\newblock \bibinfo{journal}{Human Factors} \bibinfo{volume}{27}
  (\bibinfo{year}{1985}) \bibinfo{pages}{157--162}.
\bibitem[{Troiano et~al.(2008)Troiano, Birtolo, Armenise, and
  Cirillo}]{Troiano2008242}
\bibinfo{author}{L.~Troiano}, \bibinfo{author}{C.~Birtolo},
  \bibinfo{author}{R.~Armenise}, \bibinfo{author}{G.~Cirillo},
\newblock \bibinfo{title}{Optimization of menu layouts by means of genetic
  algorithms},
\newblock \bibinfo{journal}{Lecture Notes in Computer Science (including
  subseries Lecture Notes in Artificial Intelligence and Lecture Notes in
  Bioinformatics)} \bibinfo{volume}{4972 LNCS} (\bibinfo{year}{2008})
  \bibinfo{pages}{242--253}. \DOIprefix\doi{10.1007/978-3-540-78604-7_21},
  \bibinfo{note}{conference of 8th European Conference on Evolutionary
  Computation in Combinatorial Optimization, EvoCOP 2008 ; Conference Date: 26
  March 2008 Through 28 March 2008; Conference Code:72583}.
\bibitem[{Troiano and Birtolo(2014)}]{Troiano2014433}
\bibinfo{author}{L.~Troiano}, \bibinfo{author}{C.~Birtolo},
\newblock \bibinfo{title}{Genetic algorithms supporting generative design of
  user interfaces: Examples},
\newblock \bibinfo{journal}{Information Sciences} \bibinfo{volume}{259}
  (\bibinfo{year}{2014}) \bibinfo{pages}{433--451}. \URLprefix
  \url{https://www.scopus.com/inward/record.uri?eid=2-s2.0-84889645955&doi=10.1016%2fj.ins.2012.01.006&partnerID=40&md5=b98a05d360090523513f5eb34b2e10b9}.
  \DOIprefix\doi{10.1016/j.ins.2012.01.006}.
\bibitem[{Troiano et~al.(2016)Troiano, Birtolo, and
  Armenise}]{troiano2016searching}
\bibinfo{author}{L.~Troiano}, \bibinfo{author}{C.~Birtolo},
  \bibinfo{author}{R.~Armenise},
\newblock \bibinfo{title}{Searching optimal menu layouts by linear genetic
  programming},
\newblock \bibinfo{journal}{Journal of ambient intelligence and humanized
  computing} \bibinfo{volume}{7} (\bibinfo{year}{2016})
  \bibinfo{pages}{239--256}.
\bibitem[{Matsui and Yamada(2008)}]{matsui2008optimizing}
\bibinfo{author}{S.~Matsui}, \bibinfo{author}{S.~Yamada},
\newblock \bibinfo{title}{Optimizing hierarchical menus by genetic algorithm
  and simulated annealing},
\newblock in: \bibinfo{booktitle}{Proceedings of the 10th annual conference on
  Genetic and evolutionary computation}, \bibinfo{organization}{ACM},
  \bibinfo{year}{2008}, pp. \bibinfo{pages}{1587--1594}.
\bibitem[{Golovine et~al.(2010)Golovine, McCall, and
  Holt}]{golovine2010evolving}
\bibinfo{author}{J.-C. Golovine}, \bibinfo{author}{J.~McCall},
  \bibinfo{author}{P.~O. Holt},
\newblock \bibinfo{title}{Evolving interface designs to minimize user task
  times as simulated in a cognitive architecture},
\newblock in: \bibinfo{booktitle}{Evolutionary Computation (CEC), 2010 IEEE
  Congress on}, \bibinfo{organization}{IEEE}, \bibinfo{year}{2010}, pp.
  \bibinfo{pages}{1--7}.
\bibitem[{Goubko and Danilenko(2012)}]{goubko2012mathematical}
\bibinfo{author}{M.~Goubko}, \bibinfo{author}{A.~Danilenko},
\newblock \bibinfo{title}{Mathematical model of hierarchical menu structure
  optimization},
\newblock \bibinfo{journal}{Automation and Remote Control} \bibinfo{volume}{73}
  (\bibinfo{year}{2012}) \bibinfo{pages}{1410--1423}.
\bibitem[{Karrenbauer and Oulasvirta(2014)}]{karrenbauer2014improvements}
\bibinfo{author}{A.~Karrenbauer}, \bibinfo{author}{A.~Oulasvirta},
\newblock \bibinfo{title}{Improvements to keyboard optimization with integer
  programming},
\newblock in: \bibinfo{booktitle}{Proceedings of the 27th annual ACM symposium
  on User interface software and technology}, \bibinfo{organization}{ACM},
  \bibinfo{year}{2014}, pp. \bibinfo{pages}{621--626}.
\bibitem[{Balintfy(1964)}]{balintfy1964menu}
\bibinfo{author}{J.~L. Balintfy},
\newblock \bibinfo{title}{Menu planning by computer},
\newblock \bibinfo{journal}{Communications of the ACM} \bibinfo{volume}{7}
  (\bibinfo{year}{1964}) \bibinfo{pages}{255--259}.
\bibitem[{Lancaster(1992)}]{lancaster1992history}
\bibinfo{author}{L.~M. Lancaster},
\newblock \bibinfo{title}{The history of the application of mathematical
  programming to menu planning},
\newblock \bibinfo{journal}{European Journal of Operational Research}
  \bibinfo{volume}{57} (\bibinfo{year}{1992}) \bibinfo{pages}{339--347}.
\bibitem[{Dantzig(1990)}]{dantzig1990diet}
\bibinfo{author}{G.~B. Dantzig},
\newblock \bibinfo{title}{The diet problem},
\newblock \bibinfo{journal}{Interfaces} \bibinfo{volume}{20}
  (\bibinfo{year}{1990}) \bibinfo{pages}{43--47}.
\bibitem[{Bagnall et~al.(2001)Bagnall, Rayward-Smith, and
  Whittley}]{bagnall2001next}
\bibinfo{author}{A.~J. Bagnall}, \bibinfo{author}{V.~J. Rayward-Smith},
  \bibinfo{author}{I.~M. Whittley},
\newblock \bibinfo{title}{The next release problem},
\newblock \bibinfo{journal}{Information and software technology}
  \bibinfo{volume}{43} (\bibinfo{year}{2001}) \bibinfo{pages}{883--890}.
\bibitem[{Oulasvirta et~al.(2017)Oulasvirta, Feit, L{\"a}hteenlahti, and
  Karrenbauer}]{oulasvirta2017computational}
\bibinfo{author}{A.~Oulasvirta}, \bibinfo{author}{A.~Feit},
  \bibinfo{author}{P.~L{\"a}hteenlahti}, \bibinfo{author}{A.~Karrenbauer},
\newblock \bibinfo{title}{Computational support for functionality selection in
  interaction design},
\newblock \bibinfo{journal}{ACM Transactions on Computer-Human Interaction
  (TOCHI)} \bibinfo{volume}{24} (\bibinfo{year}{2017}) \bibinfo{pages}{34}.
\bibitem[{Drira et~al.(2007)Drira, Pierreval, and Hajri-Gabouj}]{DRIRA2007255}
\bibinfo{author}{A.~Drira}, \bibinfo{author}{H.~Pierreval},
  \bibinfo{author}{S.~Hajri-Gabouj},
\newblock \bibinfo{title}{Facility layout problems: A survey},
\newblock \bibinfo{journal}{Annual Reviews in Control} \bibinfo{volume}{31}
  (\bibinfo{year}{2007}) \bibinfo{pages}{255 -- 267}.
\bibitem[{Anjos and Vieira(2017)}]{ANJOS20171}
\bibinfo{author}{M.~F. Anjos}, \bibinfo{author}{M.~V. Vieira},
\newblock \bibinfo{title}{Mathematical optimization approaches for facility
  layout problems: The state-of-the-art and future research directions},
\newblock \bibinfo{journal}{European Journal of Operational Research}
  \bibinfo{volume}{261} (\bibinfo{year}{2017}) \bibinfo{pages}{1 -- 16}.
\bibitem[{Gen and Cheng(2007)}]{MRFLPBook}
\bibinfo{author}{M.~Gen}, \bibinfo{author}{R.~Cheng}, \bibinfo{title}{Facility
  Layout Design Problems}, \bibinfo{publisher}{John Wiley \& Sons, Inc.},
  \bibinfo{year}{2007}, pp. \bibinfo{pages}{292--329}.
\bibitem[{Pirolli and Card(1999)}]{pirolli1999information}
\bibinfo{author}{P.~Pirolli}, \bibinfo{author}{S.~Card},
\newblock \bibinfo{title}{Information foraging.},
\newblock \bibinfo{journal}{Psychological review} \bibinfo{volume}{106}
  (\bibinfo{year}{1999}) \bibinfo{pages}{643}.
\bibitem[{Liu et~al.(2017)Liu, Bailly, and Howes}]{liu2017effects}
\bibinfo{author}{W.~Liu}, \bibinfo{author}{G.~Bailly},
  \bibinfo{author}{A.~Howes},
\newblock \bibinfo{title}{Effects of frequency distribution on linear menu
  performance},
\newblock in: \bibinfo{booktitle}{Proceedings of the 2017 CHI Conference on
  Human Factors in Computing Systems}, \bibinfo{organization}{ACM},
  \bibinfo{year}{2017}, pp. \bibinfo{pages}{1307--1312}.
\bibitem[{Goldberg and Deb(1991)}]{goldberg1991comparative}
\bibinfo{author}{D.~E. Goldberg}, \bibinfo{author}{K.~Deb},
\newblock \bibinfo{title}{A comparative analysis of selection schemes used in
  genetic algorithms},
\newblock in: \bibinfo{booktitle}{Foundations of genetic algorithms},
  volume~\bibinfo{volume}{1}, \bibinfo{publisher}{Elsevier},
  \bibinfo{year}{1991}, pp. \bibinfo{pages}{69--93}.
\bibitem[{Mann and Whitney(1947)}]{mann1947test}
\bibinfo{author}{H.~B. Mann}, \bibinfo{author}{D.~R. Whitney},
\newblock \bibinfo{title}{On a test of whether one of two random variables is
  stochastically larger than the other},
\newblock \bibinfo{journal}{The annals of mathematical statistics}
  (\bibinfo{year}{1947}) \bibinfo{pages}{50--60}.

\end{thebibliography}

\newpage
\section*{Appendices} \label{Appendix} \subsection{Terminology} \label{Terminology}
This section provides a glossary for the terminology used in the paper. 

\begin{table}[H]
\centering
\begin{tabular}{l|l}
\hline
\multicolumn{1}{c|}{Term used} & \multicolumn{1}{c}{Interpretation} \\ \hline
a,b  &  Fitts' law constants               \\ \hline
X  & Position of a command to a group       \\ \hline
Y  & Position of a command to a tab         \\ \hline 
Q  & Position of a group on a tab         \\ \hline 
Z  & Commands in the same group         \\ \hline 
W  & Command on the same tab         \\ \hline 
R  & Row position of a command in a tab         \\ \hline 
t  & The time required to reach a command         \\ \hline 
S  & Proceeding groups        \\ \hline
$\mathbb{S}$ & First group on each tab        \\ \hline
 $\xi$& Used groups       \\ \hline
 $\beta$ & Used tabs       \\ \hline
$\Theta$   &    Position of groups before a specific group \\ \hline
 P  & Starting position of a group within a tab       \\ \hline
 $\lambda$   &   Weight of importance of each objective function  \\ \hline
$\mathbb{A}$   & Pairwise association of commands       \\ \hline
U      &    First command of a group    \\ \hline
$\Phi$       &  Searching effort for finding a command      \\ \hline
$\alpha$       & True-positive time of finding a command       \\ \hline
 $\sigma$      &  False-positive time of finding a command           \\ \hline
$\delta$       & False-negative time of finding a command      \\ \hline
$\Omega$       & Penalty for a command if it is placed on a non-standard tab\\ \hline
$\Pi, \Xi$        &  Change in position for command $i$ from original layout to new layout      \\ \hline

\end{tabular}
\caption{Glossary of terminology used in the paper}
\end{table}

\subsection{Details of the MRF constraints} \label{MRFConstraints}
The following constraints apply to the MRF:
\begin{align}
&t_i= (a+b\log_2(\sum_{r}rR_i^r+1))+(a+b\log_2(\sum_{\tau}\tau Y_i^\tau+1)) \dots \forall\ i\in N
\end{align}
This constraint is used to calculate the selection time according to the Fitts' law. The first part on the right equation is the Fitts' law for rows and the next part is for tabs.

\begin{align}
&\left\vert{N}\right\vert\xi^c \ge \sum_{i \in N}X_i^c \dots \forall\ c\\
&\sum_{i \in N}X_i^c \ge \xi^c  \dots \forall\ c\\
&\left\vert{N}\right\vert\beta^\tau \ge \sum_{i \in N} Y_i^\tau  \dots \forall\
\tau\\
&\sum_{i \in N} Y_i^\tau \ge \beta^\tau   \dots \forall\ \tau
\end{align}
The above constraints ensure that any group or tab will be marked for
use if and only if it actually includes at least one command. 
\begin{align}
&W_{ij} \ge Z_{ij}   \dots  \forall i,j \in N\\
\intertext{This constraint ensures that if two commands are in the same group, they
share the same tab.}
&\sum_\tau Q^{c\tau} = \xi^c  \dots  \forall
c\\
&\sum_\tau Y_i^\tau = 1  \dots  \forall i \in N\\
&\sum_c X_i^c = 1  \dots  \forall i \in N
\end{align}
Every command must be placed on exactly one tab and should be
present in exactly one group. Similarly, every group (if marked for
use) should be placed on exactly one tab.
\begin{align} 
&\beta^\tau \ge Q^{c\tau}  \dots  \forall c,\tau\\
\intertext{If a group is to be placed on a specific tab, then that
tab must be marked as non-empty.} 
&\sum_r R_i^r = 1  \dots  \forall i \in N\\
\intertext{Every command must be placed on exactly one row.}
&Q^{\bar{c}\tau} \ge Q^{c\tau} + S^{c\bar{c}} + S^{\bar{c}c} - 1 \dots \forall
c, \bar{c}, \tau\\
\intertext{Two groups marked for immediate precedence must be placed on the same tab.} 
&\sum_i R_i^r \le \sum_\tau \beta^\tau   \dots  \forall r\\
\intertext{The total number of commands on any row is less than or equal to the
number of tabs being used.} 
&\sum_i R_i^r \le \sum_i R_i^{r-1}    \dots  \forall r:\ r \ge 1\\
\intertext{No intermediate rows should be left unoccupied (avoid holes).}  
&\sum_\tau \beta^\tau = \sum_c \mathbb{S}^{c}  \dots  \forall c \in C\\
\intertext{The number of tabs for use is equal to the number of groups that can be
the starting (topmost) groups.}
&\xi^c = \sum_{\bar{c} \in C} S^{\bar{c}c} + \mathbb{S}^c  \dots  \forall c \in
C\\
\intertext{If a group is used, it must either be the topmost group on its tab or be preceded by another group.} 
&R_i^r + R_j^r + W_{ij} \leq 2  \dots  \forall i,j \in N\\
\intertext{No two commands on a tab may share the same row number. We
can enforce a similar constraint for any pair of commands within a single
group.} 
&\sum_r rR_i^r \leq P^c + \sum_j X^c_j + \left\vert{N}\right\vert(1-X_i^c) \dots
\forall i \in N, \forall c \in C \label{LowerLimitOnRowNum} \\
&\sum_r rR_i^r \geq P^c - \left\vert{N}\right\vert(1-X_i^c) \dots \forall i \in
N, \forall c \in C\\
\intertext{The above constraints keep the row number for any given command within the bounds of its designated group.}
&P^{\bar{c}} \geq P^c + \sum_{i \in N}X_i^c -
\left\vert{N}\right\vert(1-\Theta^{c\bar{c}}) , \forall c,\bar{c} \in C\\
\intertext{This constraint ensures that no two groups from any tab can overlap each other.}
&X_i^c \ge X_j^c + Z_{ij} -1 \dots \forall i,j \in
N, \forall c \in C\\ 
\intertext{This constraint interconnects the $X$ variable with the $Z$ variable to ensure the logical constitution of groups. A similar constraint is enforced for the $Y$ and $W$ variables.}\notag
\end{align}

\subsection{The baseline designs for all data instances} \label{Baseline Designs for all Data Instances}
This section presents the baseline (existing) menu designs as seen in the commercial versions of Notepad, Acrobat, and Firefox applications.
\begin{figure}[!htbp] 
\centering
\scalebox{0.35}
{\includegraphics{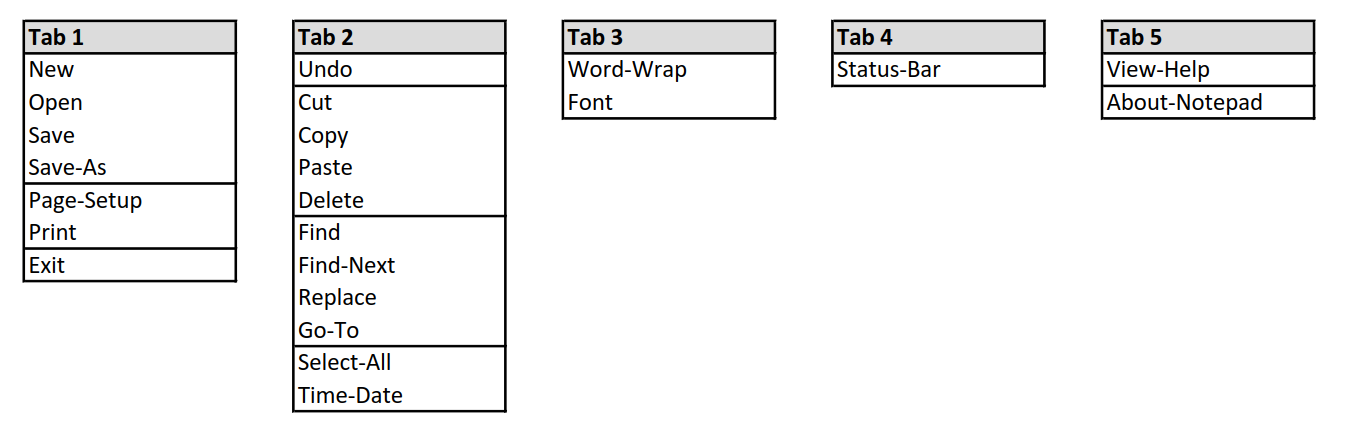}}
\caption{Notepad baseline design.}
\label{NotepadUnoptimized.PNG}
\end{figure}

\begin{figure}[!htbp] 
\centering
\scalebox{0.4}
{\includegraphics{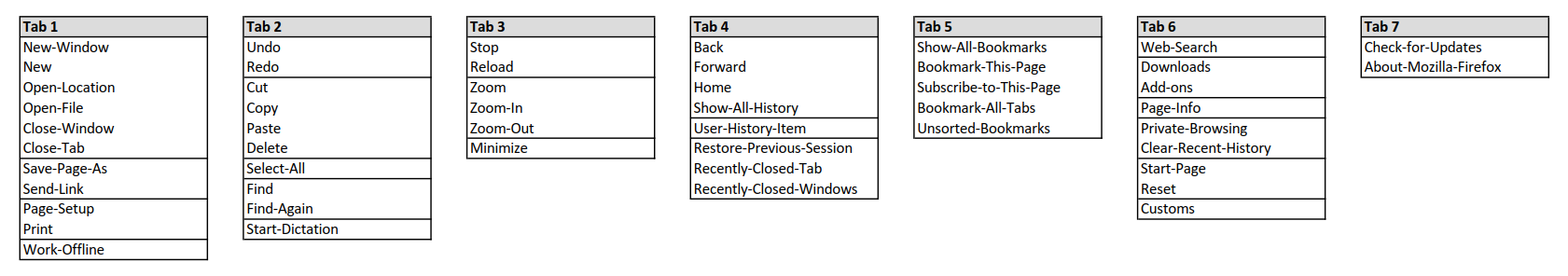}}
\caption{Mozilla Firefox baseline design.}
\label{MozillaUnoptimized.PNG}
\end{figure}

\begin{figure}[H] 
\centering
\scalebox{0.4}
{\includegraphics{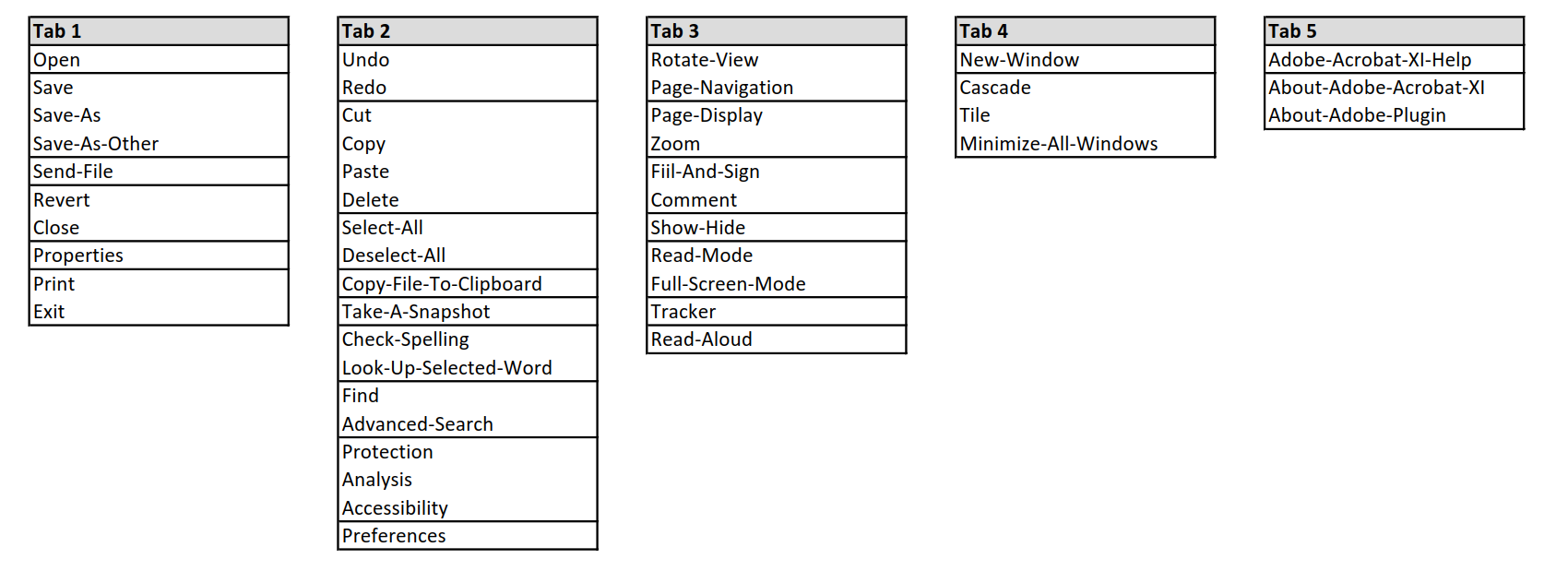}}
\caption{Adobe Reader baseline design.}
\label{AcrobatUnoptimized.PNG}
\end{figure}

\subsection{Preferred locations for specific commands} \label{Discuss Preferred Locations}
Menu systems in software applications have traditionally followed some unwritten norms regarding placement of certain key command groups in specific tabs. For example:
\begin{enumerate}
    \item Commands to open/create a new session, file or activity are predominantly located in the first (leftmost) tab of the menu.
    \item Commands to save/close an ongoing session, file or activity are predominantly located in the first (leftmost) tab of the menu.
    \item Commands to manipulate the clipboard by Cut/Copy/Paste some parts of an ongoing file or activity are never located in the first (leftmost) tab or the last (rightmost) of the menu. Rather, these commands are typically in the tab that is second from left.
    \item Commands to access 'Help' topics, to read information \textit{about} the current software application, find its version or to update that application are predominantly located in the last (rightmost) tab of the menu.
\end{enumerate}
While the four norms written above have not been formally documented in standard design guidelines, the authors posit that it is rare to find common professional software application which do not follow these norms. We postulate two reasons behind such norms: (1) Designers of some seminal software applications may have logically chosen the placement of these command groups. (2) The ingrained practice has continued unchallenged and become an essential part of user expectation. Effectively, a practice was started and no one saw any major reason to change it. 

The norms written above are extremely generic and are not restricted to any specific domain or topic. It is conceivable that specific domains, topics or business areas will have more such norms specific to practitioners of that topic. We assume that such information is available as input parameter for our menu design process. For a few (say around 5-10\% of total) commands, we assume that the preferred location specification $\mathbb{L}$ is provided in terms of the tab number where the command be preferably expected.   

\end{document}